\documentclass[a4paper,11pt]{article}

\pdfoutput=1 % if your are submitting a pdflatex (i.e. if you have
\usepackage{jheppub} % for details on the use of the package, please
\usepackage[T1]{fontenc} % if needed
\usepackage{xspace}
\usepackage{atlasphysics}
\usepackage{lineno}
\usepackage{float}
\usepackage{mathrsfs}
\usepackage{graphicx}
\usepackage{hyperref}
\usepackage{multirow}
\usepackage{rotating}
\usepackage{url}
\usepackage{textpos}
\usepackage{color}
\usepackage{epstopdf}
\usepackage{colortbl}

\usepackage{preprintcover}  % load package
\PreprintCoverPaperTitle{\boldmath Search for long-lived neutral particles decaying into lepton jets in proton--proton collisions at $\sqrt{s}$~=~8~TeV with the ATLAS detector}  % paper title
\PreprintIdNumber{CERN-PH-EP-2014-209}  % CERN preprint number
\PreprintCoverAbstract{Several models of physics beyond the Standard Model predict neutral particles that decay into final states consisting of collimated jets of light leptons and hadrons (so-called ``lepton jets''). These particles can also be long-lived with decay length comparable to, or even larger than, the LHC detectors' linear dimensions. This paper presents the results of a search for lepton jets in proton--proton collisions at the centre-of-mass energy of $\sqrt{s}$~=~8~TeV in a sample of 20.3~fb$^{-1}$ collected during 2012 with the ATLAS detector at the LHC. Limits on models predicting Higgs boson decays to neutral long-lived lepton jets are derived as a function of the particle's proper decay length.}  % paper abstract
\PreprintJournalName{JHEP}  % journal name
\newcommand{\pythia}{{\tt PYTHIA 8.165}}
\newcommand{\mcatnlo}{{\tt MC@NLO 4.06}}
\newcommand{\geant}{{\tt GEANT4}}
\newcommand{\bridge}{{\tt BRIDGE}}
\newcommand{\madgraph}{{\tt MadGraph}}

\newcommand{\mydelphi}{\Delta \phi}
\newcommand{\myabsdelphi}{$|\Delta \phi|$}
\newcommand{\mydelr}{$\Delta R$\xspace}

\newcommand{\absetaleq}{$\mid\eta\mid$~$\leq$\xspace}
\newcommand{\absetaint}{$\leq$~$\mid\eta\mid$~$\leq$\xspace}
\newcommand{\intlumi}{20.3~fb$^{-1}$\xspace}

\newcommand{\gdmumu}{\gammad$\rightarrow \mu\mu$\xspace}

\newcommand{\gdepai}{\gammad$\rightarrow  ee /\pi\pi$\xspace}
\newcommand{\higgstwogd}{$H\rightarrow2\gamma_{\rm d}+X$\xspace}
\newcommand{\higgsfourgd}{$H\rightarrow4\gamma_{\rm d}+X$\xspace}

\newcommand{\scalar}{$s_{\rm {d_{1}}}$\xspace}

\newcommand{\gdecamu}{\gammad$\rightarrow \mu \mu$\xspace}

\newcommand{\mytev}{TeV\xspace}
\newcommand{\mygev}{GeV\xspace}
\newcommand{\mymev}{MeV\xspace}

\newcommand{\myrts}{$\sqrt{s}$\xspace}
\newcommand{\myet}{$E_{\textrm{T}}$\xspace}

\newcommand{\ctau}{c$ \tau$\xspace}
\newcommand{\sigmabr}{$\sigma\times$BR\xspace}
\newcommand{\SMs}{\small SM}

\newcommand{\SumpT}{$ \Sigma p_{\rm{T}}$\xspace}
\newcommand{\maxsumpt}{max$\{\Sigma p_{\rm{T}}$\}\xspace}
\def\Zmumu{\ensuremath{Z \rightarrow \mu\mu}}
\newcommand{\fb}{$\rm fb^{-1}$\xspace}
\newcommand{\LJs}{LJs\xspace}
\newcommand{\LJ}{LJ\xspace}
\newcommand{\roi}{RoI}
\newcommand{\rois}{RoIs}
\newcommand{\lxy}{$L_{xy}$}
\newcommand{\gammad}{$\gamma_{\rm d}$\xspace}
\newcommand{\gammads}{dark photons\xspace}

\newcommand{\tagandprobe}{tag-and-probe}
\newcommand{\noprompt}{non-prompt}

\newcommand{\etaphid}{$\sqrt{(\eta_{1}-\eta_{2})^{2}+(\phi_{1}-\phi_{2})^{2}}$\xspace}
\newcommand{\etaphidik}{$\sqrt{(\eta_{i}-\eta_{k})^{2}+(\phi_{i}-\phi_{k})^{2}}$\xspace}
\title{\boldmath Search for long-lived neutral particles decaying into lepton jets in proton--proton collisions at $\sqrt{s}$~=~8~TeV with the ATLAS detector}
\author{The ATLAS Collaboration}
\abstract{Several models of physics beyond the Standard Model predict neutral particles that decay into final states consisting of collimated jets of light leptons and hadrons (so-called ``lepton jets''). These particles can also be long-lived with decay length comparable to, or even larger than, the LHC detectors' linear dimensions. This paper presents the results of a search for lepton jets in proton--proton collisions at the centre-of-mass energy of \myrts~=~8~\mytev in a sample of \intlumi collected during 2012 with the ATLAS detector at the LHC. Limits on models predicting Higgs boson decays to neutral long-lived lepton jets are derived as a function of the particle's proper decay length.}
\begin{document}
\maketitle
\flushbottom
%
%%%%%%%%%%%%%%%%%%%%%%%%% S E C T I O N %%%%%%%%%%%%%%%%%%%%%%%%%%%%%%%%%%%%%%%%%
\section{Introduction}
\label{sec:intro}
Several possible extensions of the Standard Model (SM) predict the existence of a hidden sector that is weakly coupled to the visible one (e.g. refs.~\cite{b1,b2,b3,b4,b5,b10b}). Depending on the structure of the hidden sector and its coupling to the SM, some unstable hidden states may be produced at colliders and decay back to SM particles with sizeable branching fractions. For example, in supersymmetric theories, the lightest visible super-partner may decay into hidden particles, some of which can decay back to the visible sector (see e.g. refs.~\cite{b2,b10b,b11new}). Several other distinct, non-supersymmetric, examples exist (see e.g. refs.~\cite{b1,b3,b4,b5}). If the lightest unstable hidden states have masses in the \mymev to \mygev range, they would decay mainly to leptons and possibly light mesons. \\
An extensively studied case is one in which the two sectors couple via the vector portal, in which a light hidden photon (dark photon, \gammad) mixes kinetically with the SM photon. If the hidden photon is the lightest state in the hidden sector, it decays back to SM particles with branching fractions that depend on its mass \cite{b10b, volanskynew1, epsilon}.
For the case in which the \gammad kinetically mixes with hypercharge, one finds that $\epsilon$, the kinetic mixing parameter, controls both the \gammad decay branching fractions and lifetime. More generally, however, the branching fractions and lifetime are model-dependent and may depend on additional parameters. \\
Due to their small mass, these particles are typically produced with a large boost and, due to their weak interactions, can have non-negligible lifetime. As a result one may expect, from dark photon decays, collimated jet-like structures containing pairs of electrons and/or muons and/or charged pions (``lepton jets'', \LJs) that can be produced far from the primary interaction vertex of the event (displaced \LJs). \\
Neutral particles which decay far from the interaction point into collimated final states represent a challenge both for the trigger and for the reconstruction capabilities of the LHC detectors. Collimated charged particles in the final state can be difficult to disentangle due to the limited granularity of the detector. Moreover, in the absence of information from the inner tracking system, it is necessary to use the muon spectrometer (MS) for the reconstruction of tracks which originate from a secondary decay far from the primary interaction vertex (IP). \\
The high-resolution, high-granularity measurement capability of the ATLAS ``air-core'' MS is ideal for this type of search. In addition, the ATLAS inner tracking system can be used to define isolation criteria to significantly reduce, for decay vertices far from the interaction point, the otherwise overwhelming SM background from proton--proton collisions. \\
The search for displaced \LJs presented in this paper employs the full dataset collected by ATLAS during the 2012 run at \myrts~=~8~\mytev, corresponding to an integrated luminosity of \intlumi. Related searches for prompt \LJs have been performed both at the Tevatron~\cite{tevatron1,tevatron2} and at the LHC~\cite{CMS,Chatrchyan:2012cg,Aad:2012qua,muonLJ}. Additional constraints on scenarios with hidden photons are extracted from, e.g.,  beam-dump and fixed-target experiments~\cite{117,118,119,125,126,127,128,Orsay,U70,CHARM,LSND,116}, $e^{+}e^{-}$ colliders~\cite{123,124,Hades}, B-factories~\cite{129,132}, electron and muon anomalous magnetic moment measurements~\cite{120,121,122} and astrophysical observations~\cite{130,131}. \\
The properties of the \LJ, such as its shape and particle multiplicity, strongly depend on the unknown structure of the hidden sector and its couplings to the visible sector. Therefore the search criteria must be as model-independent as possible, targeting the basic experimental signatures that correspond to these objects. A mapping of the results of such a search onto a specific model can then follow. \\
After a brief description of the ATLAS detector in section~\ref{sec:ATLAS}, two simplified models of non-SM Higgs boson decays to \LJs~\cite{b10, b10b} are presented in section~\ref{sec:LJmodels}. The \LJ definition and search criteria are given in section~\ref{sec:LeptonJet}. Section~\ref{sec:EvSel_and_Bkg} deals with the \LJ search in the data collected in 2012 and with the background evaluation. It is important to test the performance of these search criteria on some models predicting the production of final states containing \LJs; the expected signal from the two models described in section~\ref{sec:LJmodels} are presented in  section~\ref{sec:LJresult}.  Systematic uncertainties are given in section~\ref{sec:Syst}. The final results of the search and their contribution to the parameter space exclusion plot for dark photons are presented in section~\ref{sec:results}. Section \ref{sec:Conclusions} summarizes the results.
%
%%%%%%%%%%%%%%%%%%%%%%%%% S E C T I O N %%%%%%%%%%%%%%%%%%%%%%%%%%%%%%%%%%%%%%%%%
\section{The ATLAS detector}
\label{sec:ATLAS}
ATLAS is a multi-purpose detector~\cite{ATLASTDR} at the LHC, consisting of an inner tracking system (ID) contained in a superconducting solenoid, which provides a 2~T magnetic field parallel to the beam direction, electromagnetic and hadronic calorimeters (EMCAL and HCAL) and a muon spectrometer (MS) that has a system of three large air-core toroid magnets. \\
The ID combines high-resolution detectors at the inner radii with continuous tracking elements at the outer radii. It provides measurements of charged particle momenta in the region of pseudorapidity \mbox{\absetaleq~2.5}.\footnote{ATLAS uses a right-handed coordinate system with its origin at the nominal interaction point in the centre of the detector and the $z$-axis coinciding with the beam pipe axis. The $x$-axis points from the interaction point to the centre of the LHC ring, and the $y$-axis points upward. Cylindrical coordinates ($r$,$\phi$) are used in the transverse plane, $\phi$ being the azimuthal angle around the beam pipe. The pseudorapidity is defined in terms of the polar angle $\theta$ as \mbox{$\eta$ = $-$ln tan($\theta$/2)}.} The highest granularity is obtained around the vertex region using semiconductor pixel detectors arranged in three barrels at average radii of 5~cm, 9~cm, and 12~cm, and three disks on each side, between radii of 9~cm and 15~cm, followed by four layers of silicon microstrip detectors and by a transition radiation tracker.
The electromagnetic and hadronic calorimeter system covers \mbox{\absetaleq~4.9} and, at \mbox{$\eta$~=~0}, has a total depth of 9.7 interaction lengths (22 radiation lengths in the electromagnetic part). The MS provides trigger information (\mbox{\absetaleq~2.4}) and momentum measurements (\mbox{\absetaleq~2.7}) for charged particles entering the muon spectrometer. It consists of one barrel (\mbox{\absetaleq~1.05}) and two endcaps (1.05 \mbox{\absetaint~2.7}), each with 16 sectors in $\phi$, equipped with fast detectors for triggering and with chambers measuring the tracks of the outgoing muons with high spatial precision. The MS detectors are arranged in three stations at increasing distance from the IP: inner, middle and outer. Monitored drift tubes are used for precision tracking in the region \mbox{\absetaleq~2.7}, except for the innermost layer which uses cathode strip chambers in the interval 2.0 $\leq$ \mbox{\absetaleq~2.7}. The toroidal magnetic field allows for precise reconstruction of charged-particle tracks independent of the ID information. \\
The trigger system has three levels~\cite{L1TRIG} called Level-1 (L1), Level-2 (L2) and the Event Filter (EF). L1 is a hardware-based system using information from the calorimeter and MS. It defines one or more region-of-interest (\rois), geometrical regions of the detector, identified by ($\eta$, $\phi$) coordinates, containing interesting physics objects. L2 and the EF (globally called the High-Level Trigger, HLT) are software-based systems and can access information from all sub-detectors. The three planes of MS trigger chambers (resistive plate chambers in the barrel and thin gap chambers in the endcaps) are located in the middle and outer (only in the barrel) stations. The L1 muon trigger requires hits in the middle stations to create a low transverse momentum ($p_{\mathrm{T}}$) muon \roi~or hits in both the middle and outer stations for a high $p_{\mathrm{T}}$~muon \roi. The muon \rois~have a spatial extent of \mbox{0.2$\times$0.2} (\mbox{$\Delta\eta\times\Delta\phi$}) in the barrel and of \mbox{0.1$\times$0.1} in the endcaps. L1 \roi~information seeds  the reconstruction of muon momenta by the HLT, which uses precision chamber information to obtain sharper trigger thresholds.
%
%%%%%%%%%%%%%%%%%%%%%%%%% S E C T I O N %%%%%%%%%%%%%%%%%%%%%%%%%%%%%%%%%%%%%%%%%
\section{Lepton-jet models}
\label{sec:LJmodels}
\begin{figure}[ht!]
\centering
\includegraphics[width=65mm]{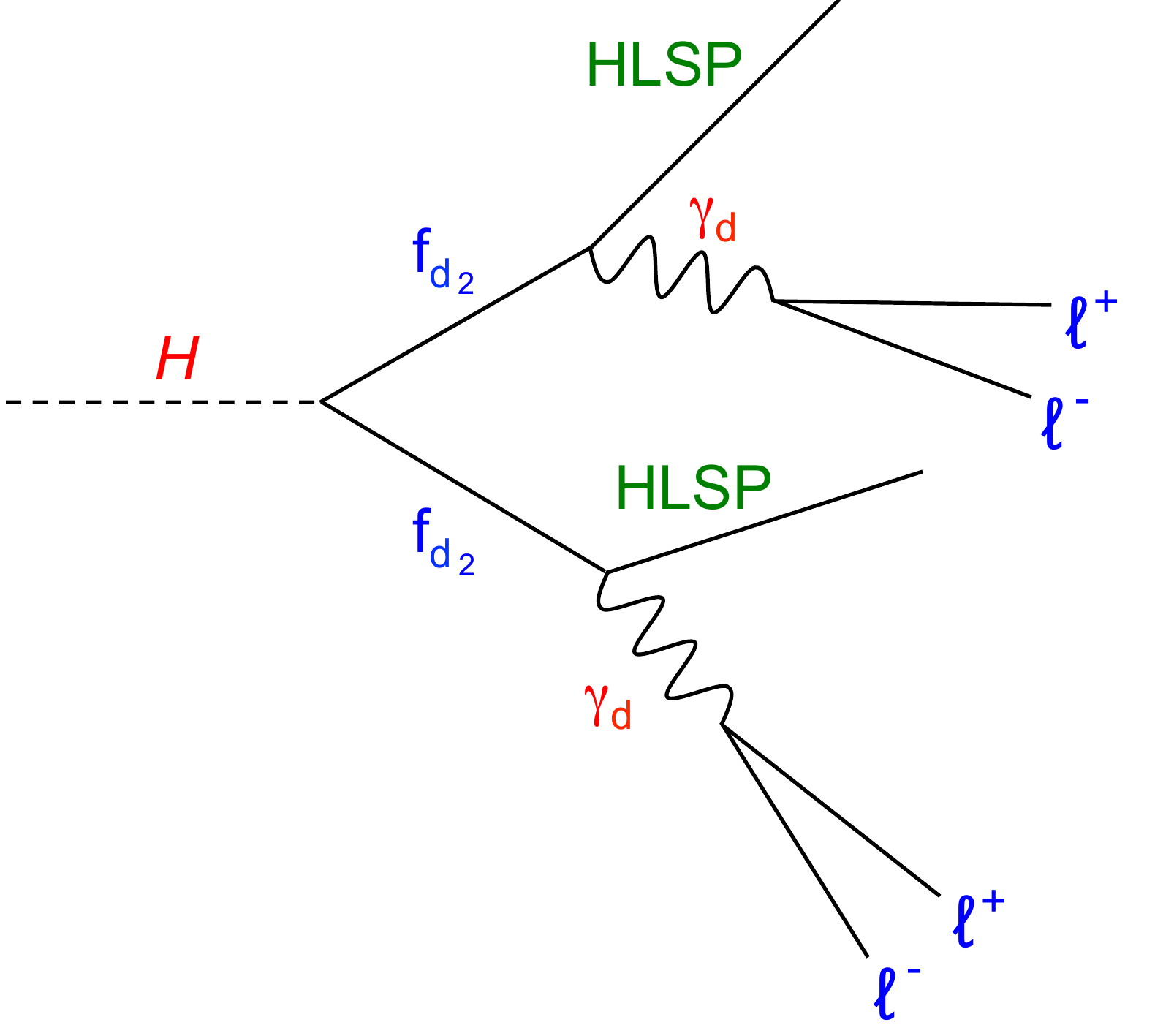}
\includegraphics[width=80mm]{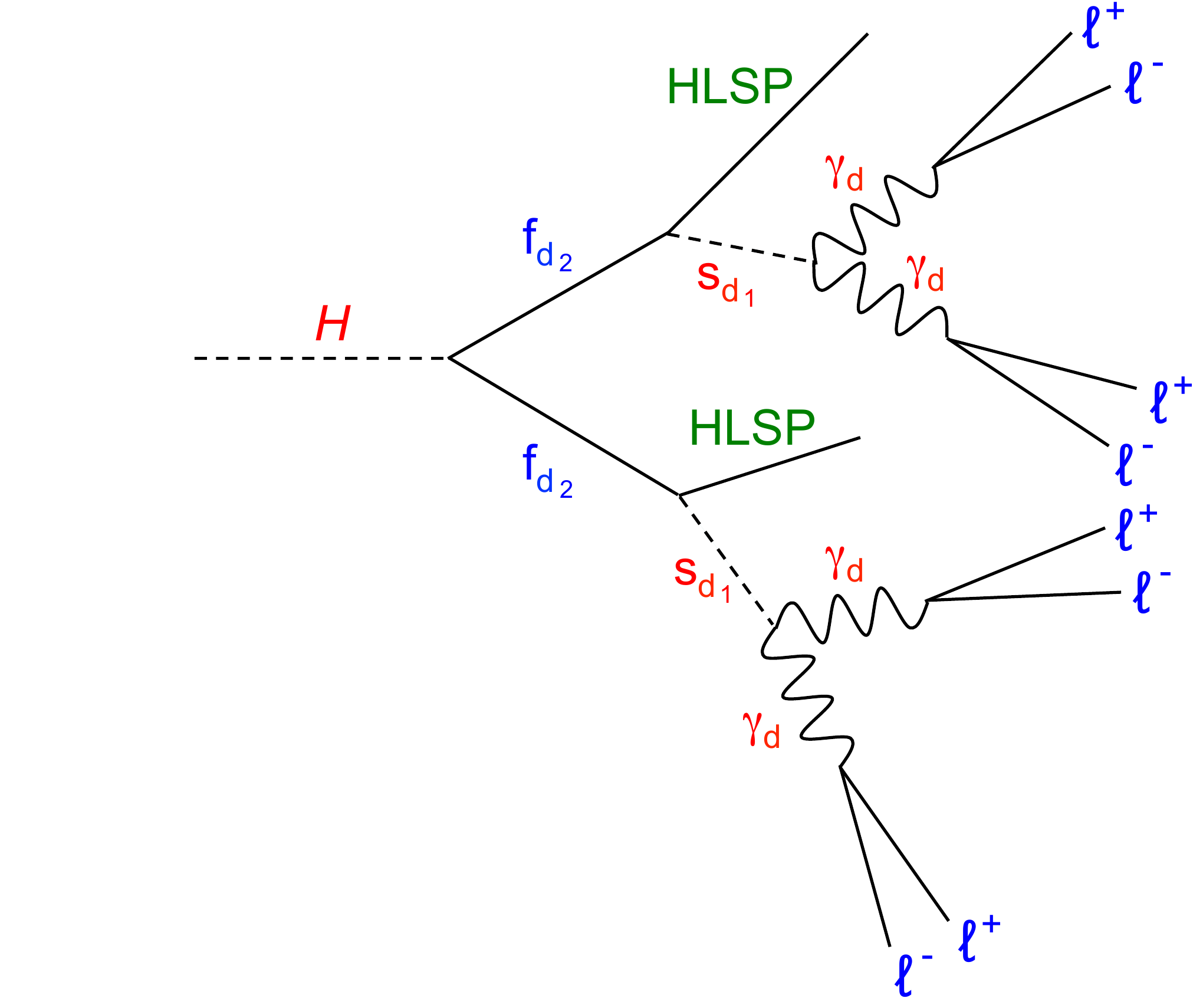}
\caption{Diagrams of the two FRVZ models used as benchmarks in the analysis. $\ell^{+}$ $\ell^{-}$ corresponds to electron/muon/pion pair decay in the final state.}
\label{fig:models}
\end{figure}

It is important to evaluate the performance of the \LJ search criteria by setting limits on models that predict LJs in the final state. Of particular relevance are models which predict non-SM Higgs boson decays to \LJs.  Indeed, the phenomenology of the Higgs boson is extremely susceptible to new couplings, and new decay channels may thus easily exist. Since the structure of the unknown hidden sector may greatly influence the properties of the \LJ, a simplified-model approach is highly beneficial. The two Falkowski--Ruderman--Volansky--Zupan (FRVZ) models~\cite{b10, b10b}, which predict non-SM Higgs boson decays to \LJs are considered. Figure~\ref{fig:models} shows diagrams for the decay of the Higgs boson to \LJs in the two models. The Higgs boson, {\it H}, decays to pairs of hidden fermions, $f_{\rm d_{2}}$. In the first model (left in figure~\ref{fig:models}) $f_{\rm d_{2}}$ decays to a dark photon, \gammad, and to a lighter hidden fermion, HLSP (Hidden Lightest Stable Particle). In the second model  (right in figure~\ref{fig:models}) $f_{\rm d_{2}}$ decays to a HLSP and to a hidden scalar, $s_{\rm d_{1}}$ that in turn decays to pairs of \gammads. For the \gammad decays, only electron, muon and pion final states are considered. In general, radiation in the hidden sector may occur, resulting in additional hidden photons. The number of such radiated photons, however, varies on an event-by-event basis and depends on unknown model-dependent parameters such as the hidden gauge coupling $\alpha_d$.\footnote{See equation 3.1 in ref.~\cite{alphad}} Therefore such a possibility is not considered here.
%
%%%%%%%%%%%%%%%%%%%%%%%%% S E C T I O N %%%%%%%%%%%%%%%%%%%%%%%%%%%%%%%%%%%%%%%%%
\section{Lepton-jet search}
\label{sec:LeptonJet}
There are a large number of possible \LJ topologies resulting from different possible hidden sectors. For instance, the \LJ shape is controlled, in part, by the typical boost of the hidden particles, which in turn is determined by the ratio of the decaying visible-sector particle's mass to the produced hidden-sector particle's mass. Additional dependence may arise from the strength of interactions within the hidden sector. For example, strong dynamics may result in broader jets such as those produced in QCD processes. Such dynamics further determines the multiplicity of particles within an \LJ. Indeed, quite generally, hidden cascade decays and possible showering may result in very dense \LJs. \\
The search presented in this paper adopts a simplified approach with a generic definition of  \LJ in order to make the analysis as model-independent as possible. An \LJ containing only one or two dark photons is considered in the optimization of the selection criteria but the search is also sensitive to more complex final states even if with lower detection efficiencies. Only displaced \LJ from \gammad decay far from the interaction point are searched for. \\
In order to characterize the ATLAS detector response to different types of displaced \LJs, an LJ gun Monte Carlo generator (MC) was developed. This MC generator is able to simulate \noprompt~\LJs produced by the decay of one \gammad or by the decay of a hidden scalar \scalar into two dark photons according to the model in ref.~\cite{b10b}. The branching ratio to electrons, muons and pions is also set according to ref.~\cite{b10b}. The \gammad  lifetime is chosen so that a large fraction of the decays occur inside the sensitive ATLAS detector volume.\footnote{The sensitive ATLAS detector volume is specified in section~\ref{sec:MC-FRVZ}.} \\
Several \LJ gun MC samples that span a wide range of the \LJ parameter space were generated. These samples are used to evaluate a suitable set of \LJ selection criteria and estimate the corresponding detection efficiency in ATLAS. For \LJ with only one \gammad the \gammad masses of 0.05, 0.15, 0.4, 0.9 and 1.5~\mygev were generated. For \LJ with two \gammads the $s_{\rm d1}$ masses of 1, 2, 5 and 10~\mygev are used. For each mass of the \scalar, only the subset of the \gammad masses kinematically allowed were generated. In order to cover a wide interval of possible \LJ production kinematics, the $p_{\mathrm{T}}$~distributions of \gammad and \scalar were taken to be uniform in the range $10 \mbox{--} 100$~\mygev and the pseudorapidity was taken to be uniform in the range from $-$2.5 to 2.5.
To study the detector response, the generated events were processed through the full ATLAS simulation chain based on \geant~\cite{GEANT4,ATLSIM}. All MC samples are simulated with pile-up interactions included and re-weighted to match the conditions of the 2012 data sample.
%
%######################## S U B - S E C T I O N ######################################
\subsection{\LJ definition}
\label{sec:LJdef}
\begin{figure*}[t!]
\centering
\includegraphics[width=48mm]{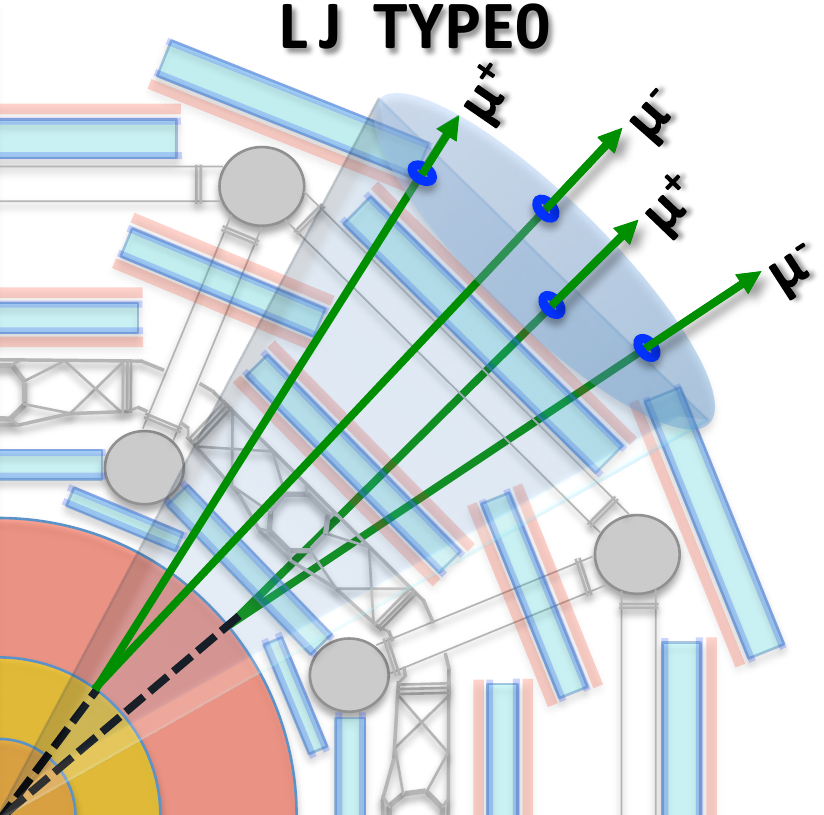}
\includegraphics[width=48mm]{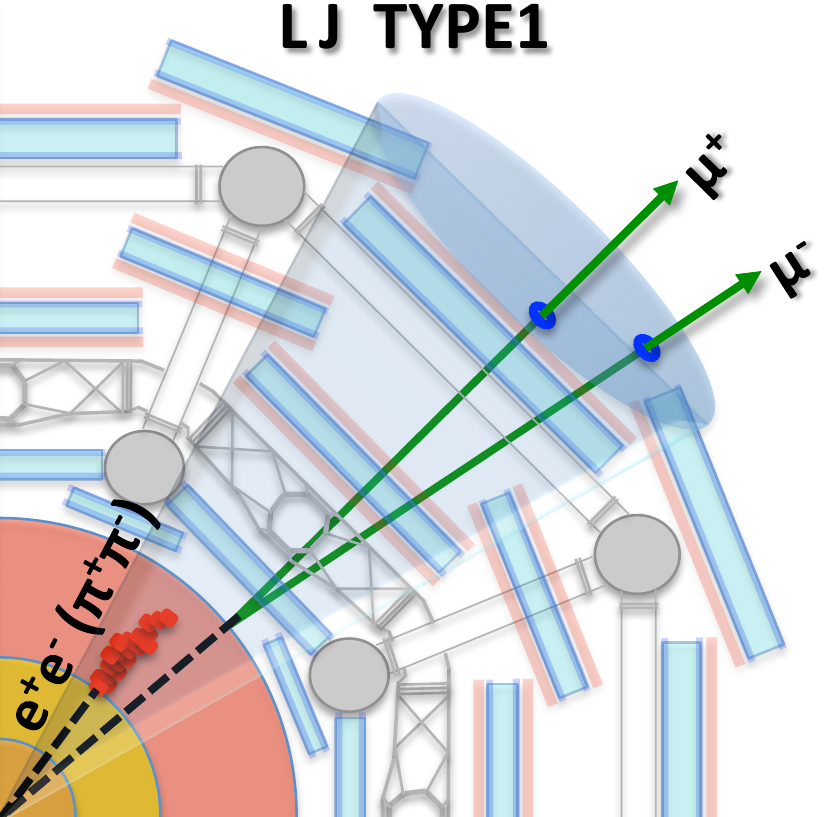}
\includegraphics[width=51mm]{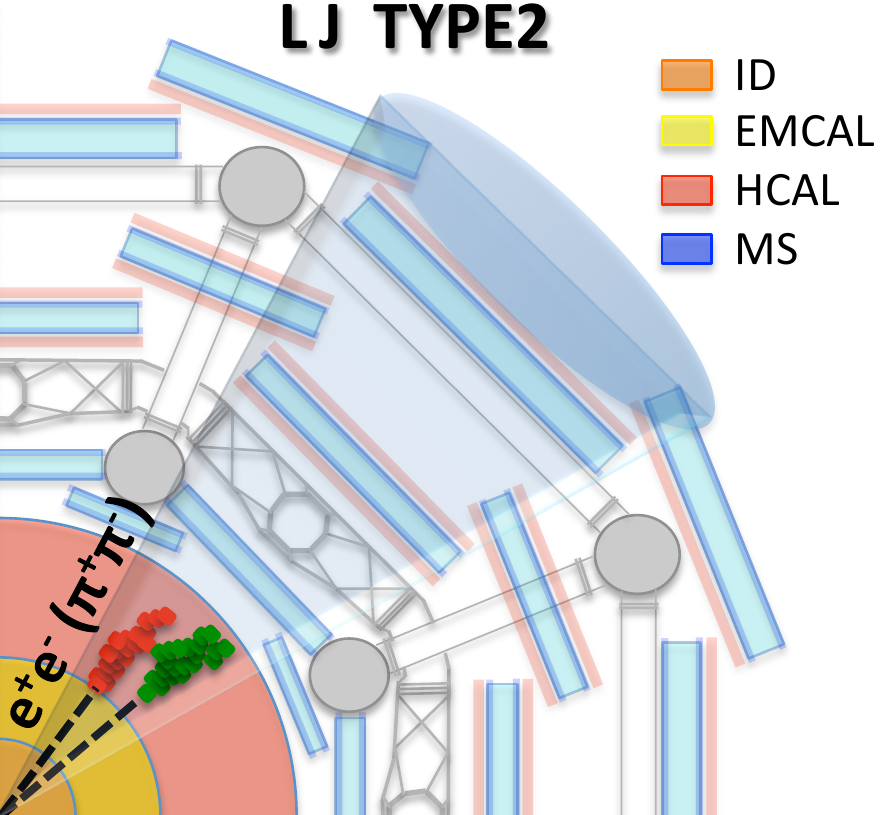}
\caption{Schematic picture of the LJ~classification according to the \gammad decay final states: left TYPE0 LJ (only muons), centre TYPE1 LJ (muons and jets), right TYPE2 LJ (only jets). LJs containing only one \gammad contribute only to TYPE0 and TYPE2.}
\label{fig:LJdef}
\end{figure*}

\begin{figure}[ht!]
\centering
\includegraphics[width=95mm]{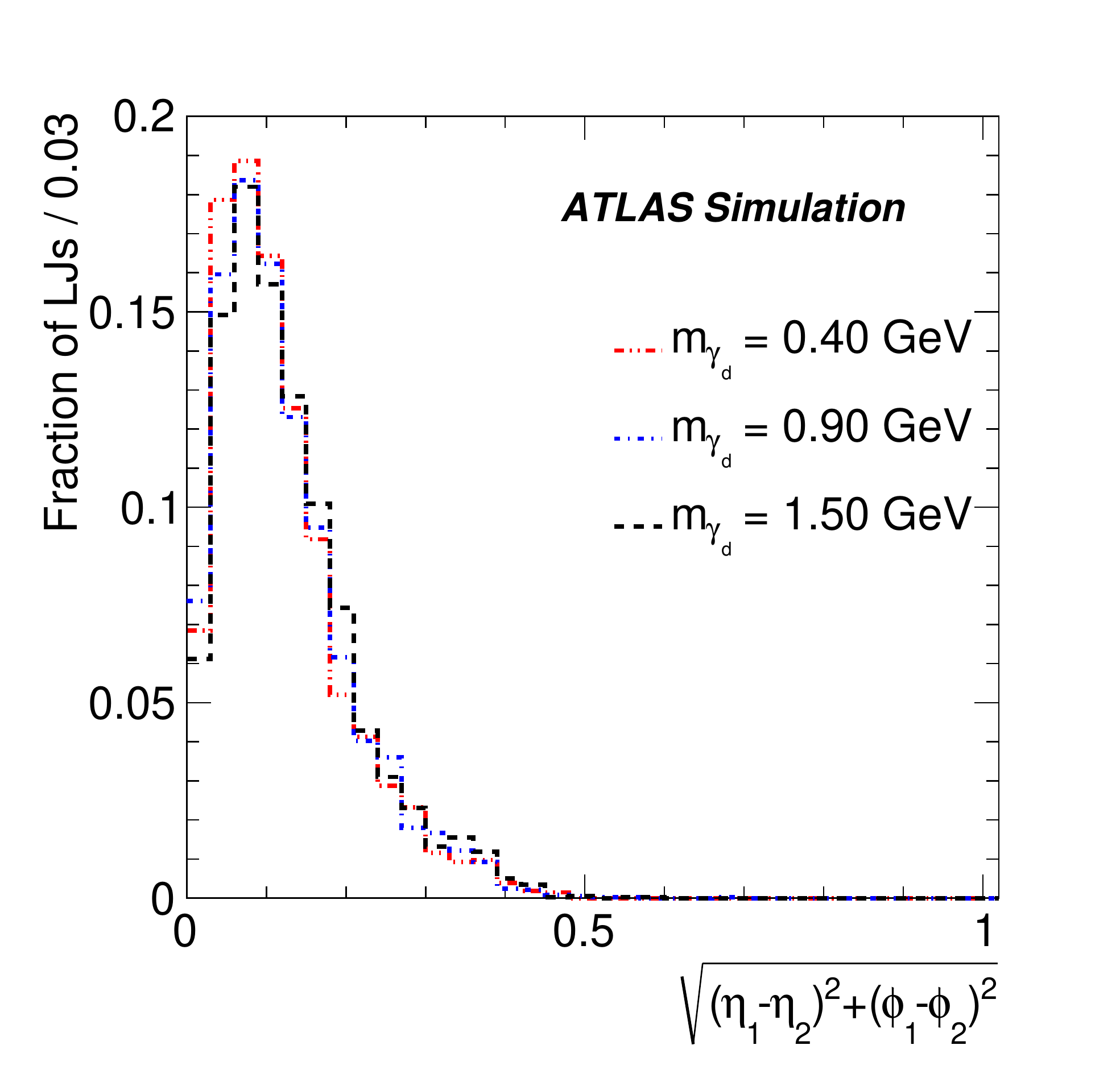}
\caption{Opening \etaphid between the two muons in an LJ produced by the decay of a single \gammad, for the simulated \gammad~mass states.}
\label{fig:maxDRfordp}
\end{figure}

\begin{figure}[ht!]
\centering
\includegraphics[width=70mm]{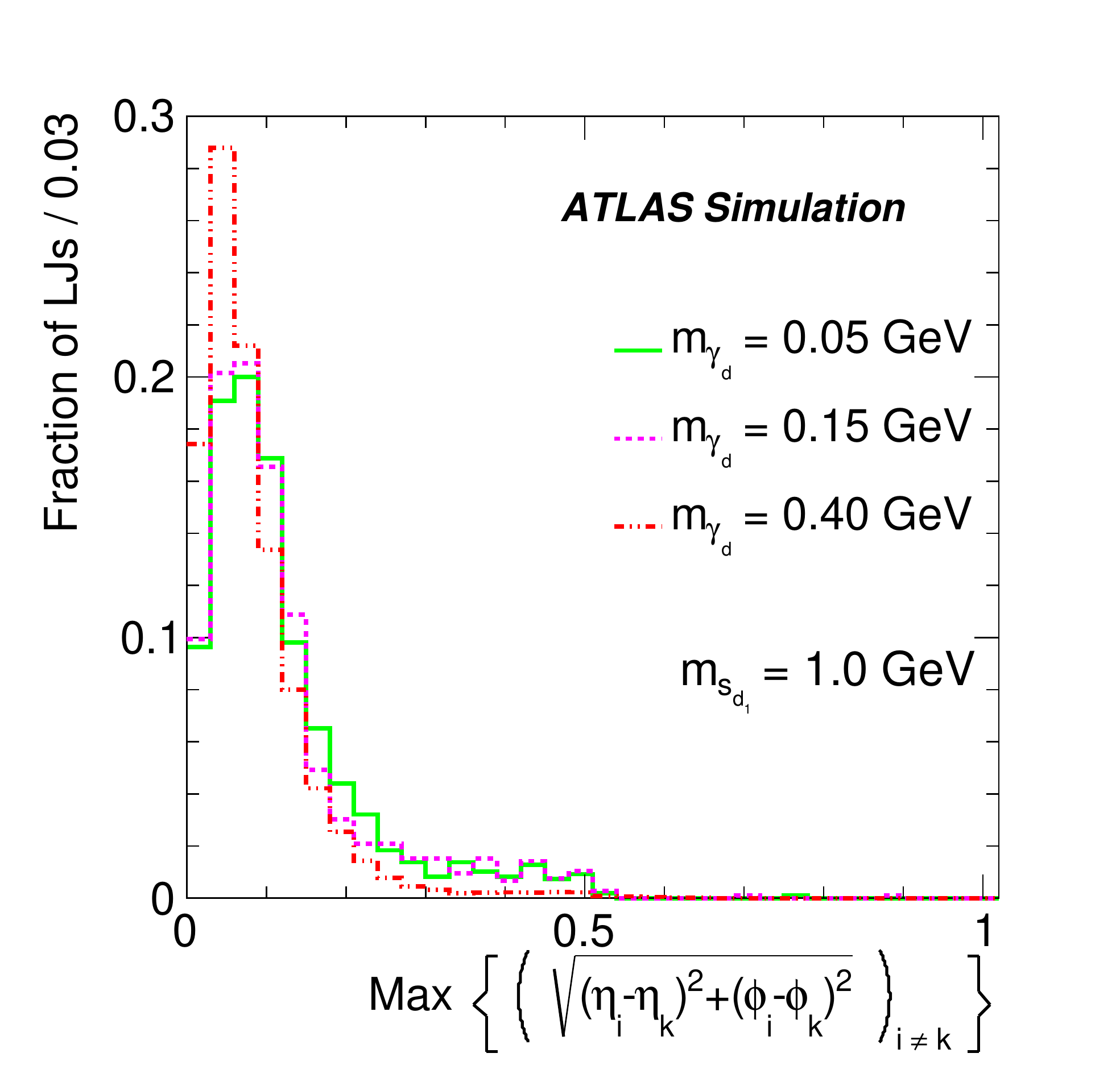}
\includegraphics[width=70mm]{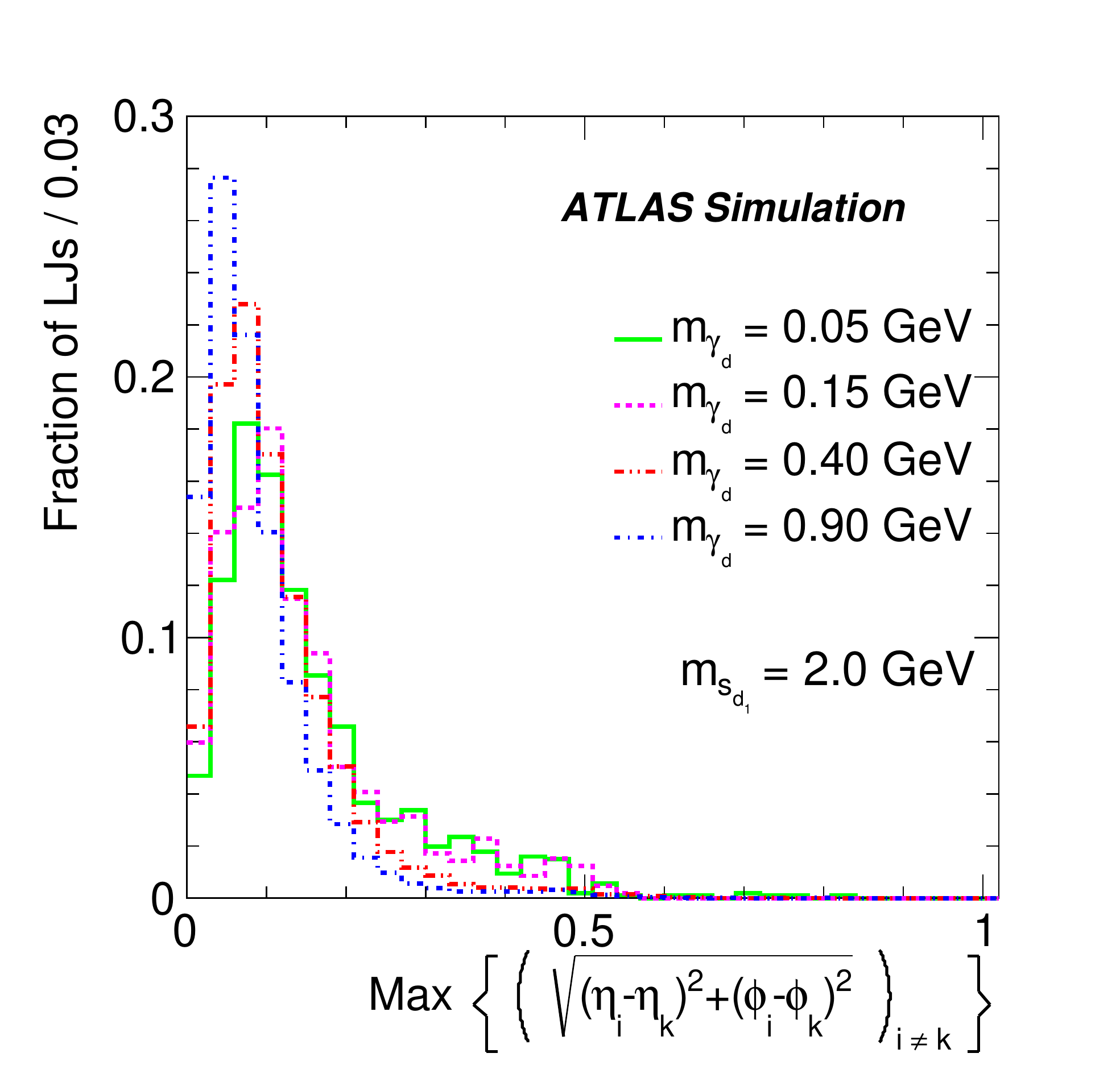}
\includegraphics[width=70mm]{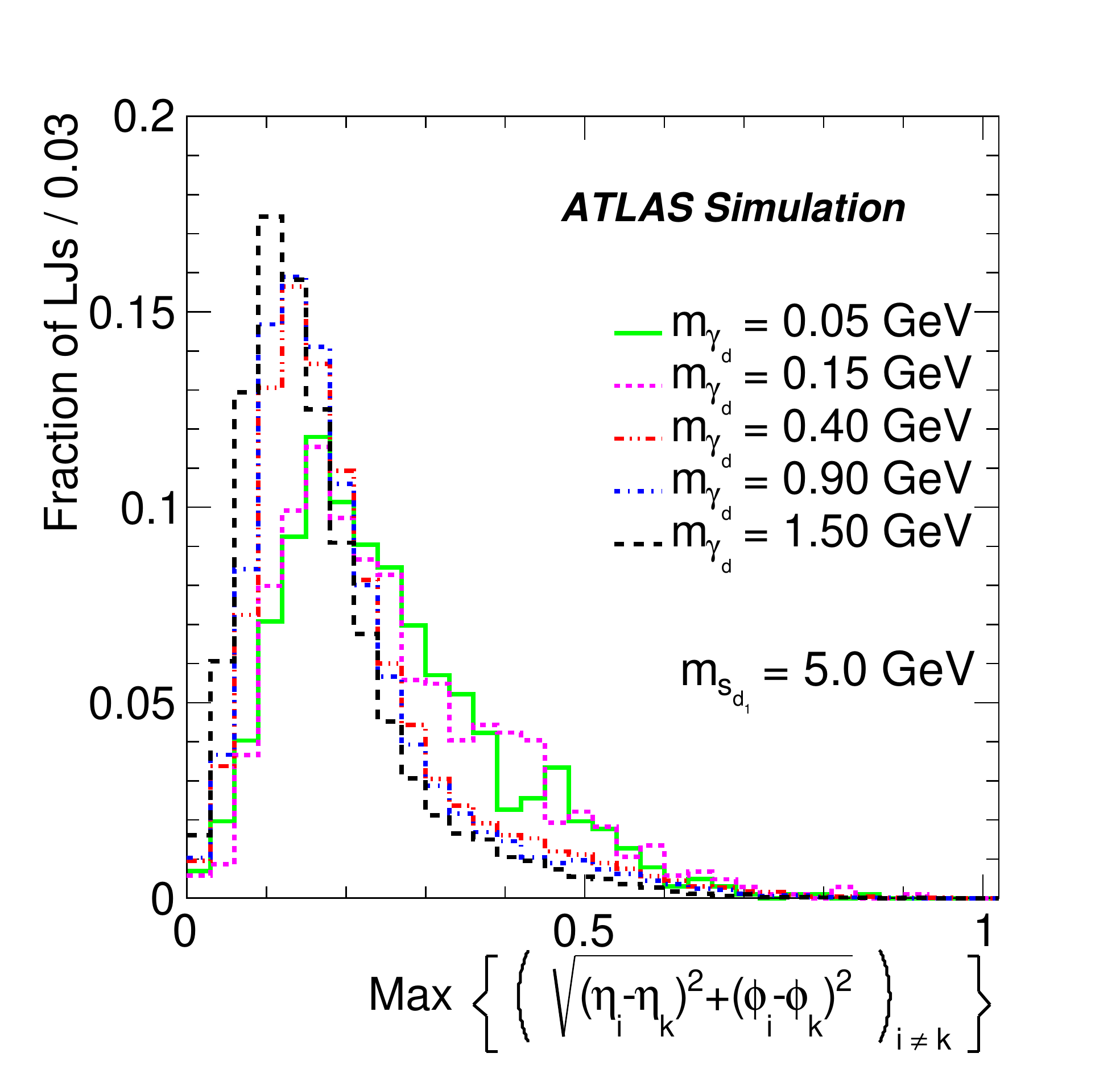}
\includegraphics[width=70mm]{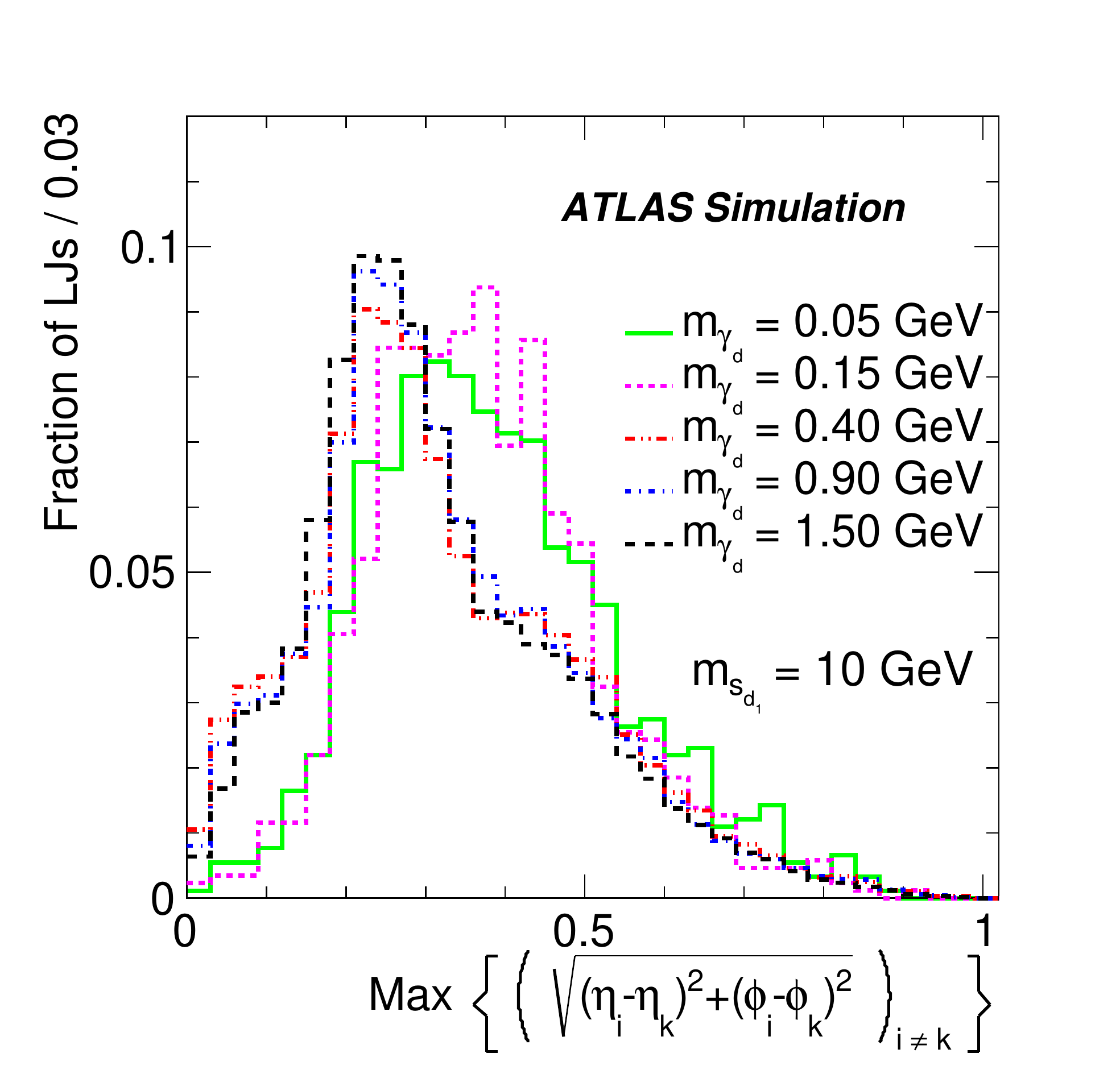}
\caption{Maximum opening \etaphidik between the reconstructed objects in the LJ (muons for TYPE0 LJ, muons and jets for the TYPE1 LJ) for the \scalar masses 1~\mygev (top left), 2~\mygev (top right), 5~\mygev (bottom left) and 10~\mygev (bottom right) and for all the kinematically allowed \gammad masses.}
\label{fig:maxDRforsd}
\end{figure}

The MC studies of the detector's response to the \LJs guide the characterization of the \LJ and the identification of variables useful for the selection of the signal. At the detector level, a \gammad decaying to a muon pair is identified by two muons in the MS, while a \gammad decaying to an electron/pion pair is seen as one or two jets in the calorimeters. A cluster of only muons and no jets in a narrow cone is the signature of an \LJ with all \gammads decaying to muon pairs. A cluster of two muons and  one or two jets is typical of an \LJ with one \gammad decaying into a muon pair and one \gammad decaying into an electron/pion pair. An \LJ with one or two \gammads, both decaying to electron/pion pairs, results in one or more jets. \\
Muons from a \gammad decay beyond the last pixel detector layer are not matched with an ID track.\footnote{The ID track reconstruction in ATLAS requires at least one hit in the pixel layers. Muon reconstruction requires a match between the muon track in the MS and an ID track (combined muons, CB).}  Therefore muon track reconstruction using only MS information (standalone muon, SA) has to be used. The search is limited to the pseudorapidity interval $-$2.5 to 2.5 corresponding to the ID coverage. \\
An anti-$k_{t}$ calorimetric jet search algorithm~\cite{AntiKt,AntiKtbis} with the radius parameter R = 0.4, is used to select \gammad decaying into an electron or pion pair. Jets must satisfy the standard ATLAS quality selection criteria~\cite{JetSelection} with the cut $p_{\mathrm{T}}$ $\geq$ 20~\mygev. The jet energy scale correction as defined in ref. \cite{JEScorrection} is applied. In the simulated \LJ gun MC samples, \LJs produced by one or two \gammads decaying to electron/pion pairs, are mostly reconstructed by the anti-$k_{t}$ algorithm as a single jet. \\
\LJs are reconstructed using a simple clustering algorithm that combines all the muons and jets lying within a cone of fixed size in ($\eta$, $\phi$) space. The algorithm is seeded by the highest-$p_{\mathrm{T}}$ muon. If at least two muons and no jets are found in the cone, the \LJ is classified as TYPE0. Otherwise, if there are at least two muons and only one jet in the cone, the \LJ found is of TYPE1. The search is then repeated with any unassociated muon until no muon seed is left. The remaining jets with electromagnetic (EM) fraction less than 0.4 and no muons in the cone are defined as TYPE2 \LJ.\footnote{EM fraction is defined as the ratio of the energy deposited in the EMCAL to the total jet energy. From the \LJ gun MC results, \gammad decaying inside the HCAL has EM fraction always below 0.4.} The \LJ line of flight is obtained from the vector sum over all muon and jet momenta in the \LJ. Figure~\ref{fig:LJdef} schematically shows the \LJ classification according to the final state.
\\
The size of the search cone for the various LJ types is optimized using the LJ gun MC samples. The cone size $\Delta R = \sqrt{(\Delta\eta)^2 + (\Delta\phi)^2}$ around the \LJ line of flight is chosen as the \mydelr that contains almost all the decay products (muons and jets) of the \gammads. Figure~\ref{fig:maxDRfordp} shows the opening angle \etaphid between the two muons for \gdmumu, with both muons reconstructed in the MS, for the three \gammad masses. Figure~\ref{fig:maxDRforsd} shows the maximum opening \etaphidik between the reconstructed objects in the TYPE0 and TYPE1 LJs, produced by the decay of two \gdmumu or one \gdmumu and one \gdepai, for various masses of the hidden scalar and of the dark photon. All these distributions show that a \mydelr~= 0.5 is adequate to contain almost all the decay products. In summary the \LJs are classified as:
\begin{itemize}
\item TYPE0 - to select \LJs with all \gammads decaying to muons. This type selects \gammad decays beyond  the pixel detector up to the first trigger plane of the MS.
\item TYPE1 - to select \LJs with one \gammad decaying to a muon pair and one \gammad decaying to an electron/pion pair. The range of decay distances targeted by TYPE1 LJ extends from the last ID pixel layer up to the end of the HCAL, for \gammad decaying into an electron/pion pair, and from the last ID pixel layer up to the first trigger plane of the MS, for the \gammad decays to muons.
\item TYPE2 - to select \LJs with all \gammads decaying to electron/pion pairs in the HCAL. The requirement of low EM fraction is necessary in order to reduce the overwhelming background due to SM multi-jet production.
\end{itemize}
The variables and the relative requirements useful for the background rejection of the individual \LJ are discussed in  section~\ref{sec:LJbkg}.
%
%######################## S U B - S E C T I O N ######################################
\subsection{LJ selection and background rejection}
\label{sec:LJbkg}
\begin{figure}[t!]
\centering
\includegraphics[width=95mm]{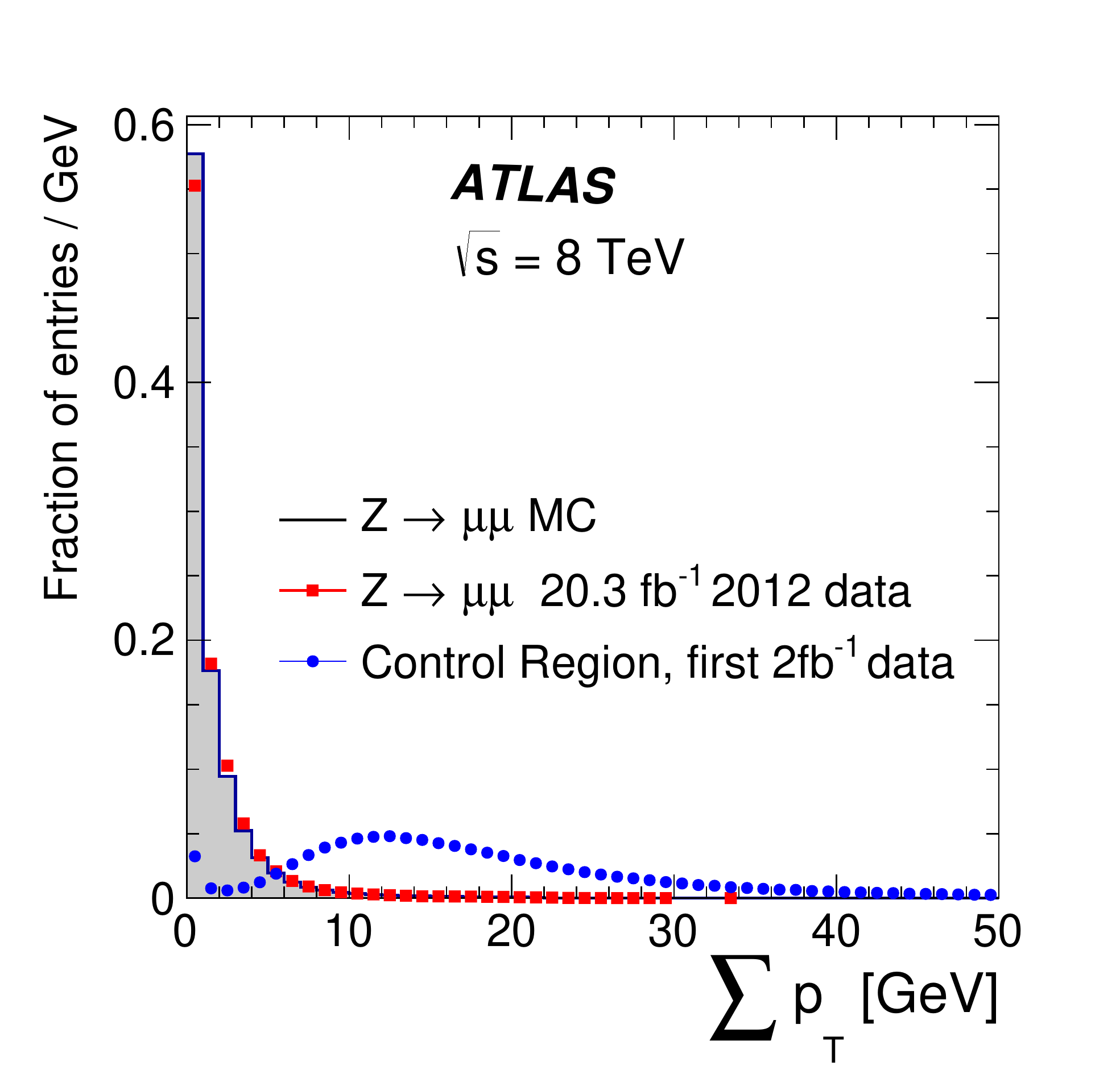}
\caption{Distributions of \SumpT : (filled dot)  control sample of the first 2~fb$^{-1}$ of 2012 data, (filled square) $Z \to \mu\mu$ in 2012 data and (solid line) $Z \to \mu\mu$ MC sample. All distributions are normalized to unit area.}
\label{fig:sumpt_cut}
\end{figure}

The main sources of background to the \LJ signal are multi-jet production and cosmic-ray muons that cross the detector in time coincidence with a bunch-crossing interaction. A sample of events collected in the empty bunch crossings is used to study the cosmic-ray background. To reduce contamination of LJ TYPE0 and TYPE1 by cosmic-ray muons, a requirement on the transverse and longitudinal impact parameters of the MS track at the primary vertex of $|d_0|~<~200$ mm and $|z_0|~<~270$ mm is used. The effect of these requirements on the \gdmumu~decay was evaluated using the \LJ gun MC of single \gammad (masses 0.4, 0.9 and 1.5~\mygev), decaying to muon pairs beyond the last pixel layer. The expected signal is reduced by about $10\mbox{--}15\%$ for decays in the ID, $15\mbox{--}25\%$ for decays in the calorimeter system and $25\mbox{--}50\%$ for decays in the MS, while the cosmic-ray background is reduced by a factor of about $200$. Since this search looks for \noprompt~\LJs, the requirement that muon tracks have no matched track in the ID (not-combined muons, NC) for TYPE0 and TYPE1 \LJs removes about 80$\%$ of the background coming from processes with production of prompt and quasi-prompt muons.\footnote{The ID efficiency for prompt or quasi-prompt muons is greater than 99$\%$ \cite{ATLASTDR}.} \\
Energy deposits in the calorimeter due to cosmic-ray muons can be reconstructed as jets, creating a background to the TYPE1 and TYPE2 \LJ selections. The variable used to remove jets from background cosmic-ray events is the timing, defined as the weighted mean time difference between $t~=~0$ (bunch-crossing time) and the time of energy deposition in the calorimeter cells. Rejecting jets with timing outside the interval between $-$1 ns and 5 ns removes a large fraction of the cosmic-ray jets, with a very small loss of signal. \\
The main background source for TYPE2 LJ is the production of multi-jet events. To study this background a control sample corresponding to the first 2~\fb of the 2012 data is used. The events were selected by single-jet triggers with the lowest available thresholds of 15~\mygev and 35~\mygev. The \LJ reconstruction algorithm is applied to this control sample. The requirement on the EM fraction and an additional requirement on the jet width were optimized by maximizing the signal significance (see eq. (97) of ref.~\cite{Significance}) defined as
\begin{equation}
 \rm \sqrt{ 2\cdot((s+b)\cdot\ln(1+s/b)-s)},
  \label{eq:significance}
\end{equation}
where $s$ and $b$ are the expected number of signal and background events, respectively.\footnote{The jet width $W$ is defined as:
\begin{equation}\label{eq:width}
\rm W=\frac{\sum_i \Delta R^{i}\cdot p_{\mathrm{T}}^{i}} {\sum_i p_{\mathrm{T}}^{i}},
\end{equation}
\noindent
where $\rm \Delta R^i \:$=$\rm \:\sqrt{(\mydelphi_i)^2 + (\Delta \eta_i)^2}$ is the distance between the jet axis and the $i^{th}$ jet constituent and $p_{\mathrm{T}}^{i}$ is the constituent $p_{\mathrm{T}}$ with respect to the beam axis.} The maximum significance for the EM fraction for TYPE2 LJ is obtained by requiring a jet EM fraction to be less than 0.1; this provides 99.9$\%$ multi-jet background rejection. A similar optimization leads to requiring a jet width less than 0.1 (80$\%$ multi-jet background rejection). In the transition regions between barrel and endcap calorimeters ($1.0<|\eta|<1.4$), where there is a discontinuity in the EMCAL coverage, many jets exhibit a fake low EM fraction. Removal of jets with $1.0<|\eta|<1.4$ rejects 30$\%$ of this type of background. An additional requirement of $|\eta|<2.5$ is also applied in order to have a  jet coverage consistent with that of the ID. \\
Non-prompt \LJs are expected to be highly isolated in the ID. Therefore the multi-jet background can be significantly reduced by requiring track isolation around the \LJ direction in the ID. The track isolation variable \SumpT (ID isolation) is defined as the sum of the transverse momenta of the tracks with $p_{\mathrm{T}}$~$>$ 500~\mev, reconstructed in  the ID and matched to the primary vertex of the event, inside a cone of size \mydelr~= 0.5 around the direction of the \LJ.\footnote{A requirement on the transverse and longitudinal impact parameters of the tracks at the primary vertex of $|d_0|~<~10$ mm and $|z_0|~<~10$ mm is used. The requirement of matching to the main primary interaction vertex helps in reducing the dependence of \SumpT on the pile-up events.} The primary interaction vertex is defined to be the vertex whose constituent tracks have the largest $\Sigma p_{\rm{T}}^2$. Figure~\ref{fig:sumpt_cut} shows the ID isolation distribution in the control sample of 2012 data selected by single-jet triggers. The ID isolation is validated with 2012 data using muons coming from a selected sample of $Z \to \mu\mu$ decays.\footnote{In this case the \pt~of the ID track matched to the muon is removed from the \SumpT.} The \SumpT distribution obtained from the $Z \to \mu\mu$ data sample agrees very well with the distribution obtained from the $Z \to \mu\mu$ MC sample, as shown in figure~\ref{fig:sumpt_cut}. A \SumpT$\leq$ 3~\mygev requirement removes 97$\%$ of the multi-jet background while maintaining a very high \LJ signal selection efficiency.
%
%######################## S U B - S E C T I O N ######################################
\subsection{LJ reconstruction efficiency}
\label{sec:LJeffi}
\begin{figure}[t!]
\centering
\includegraphics[width=70mm]{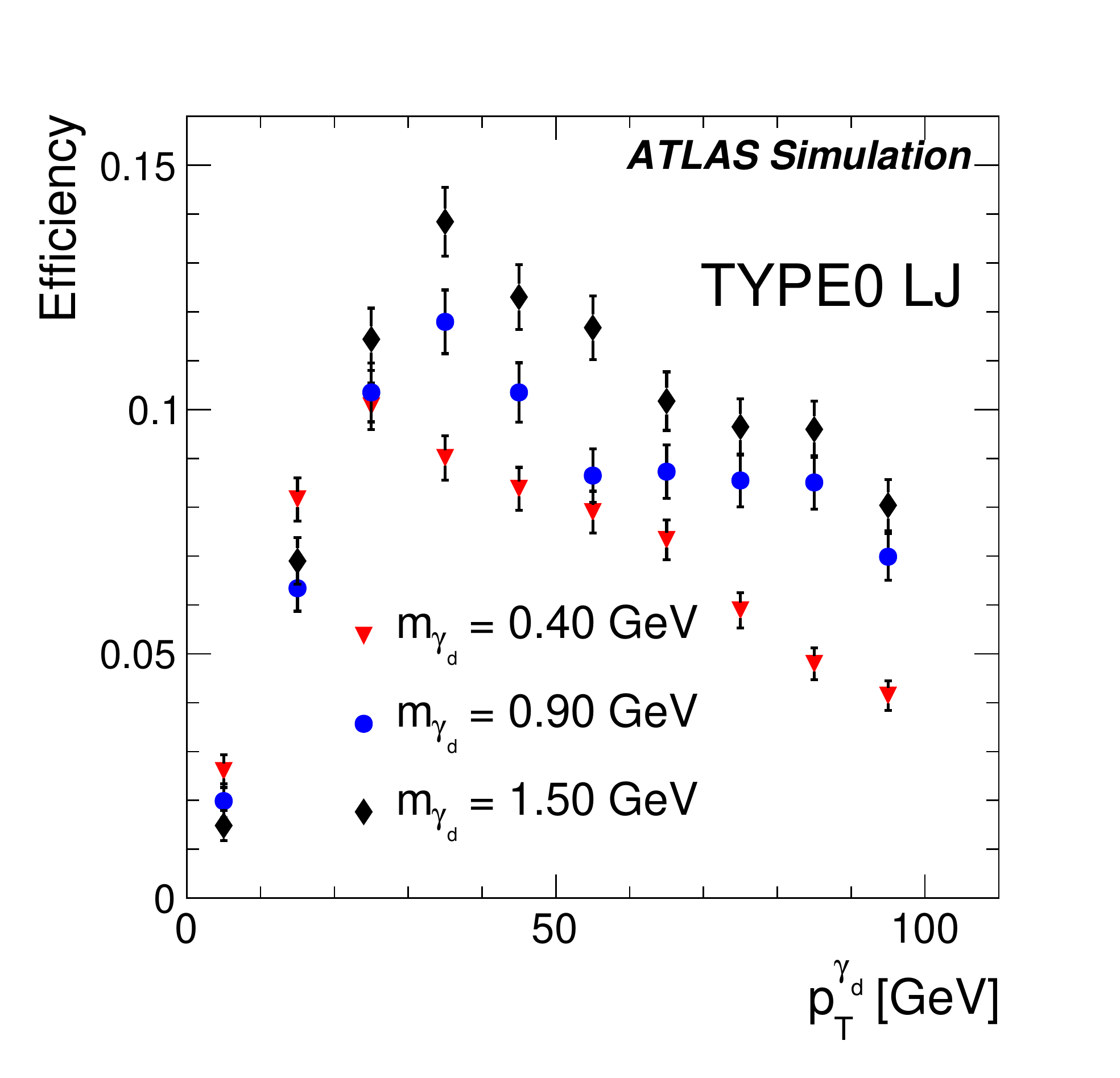}
\includegraphics[width=70mm]{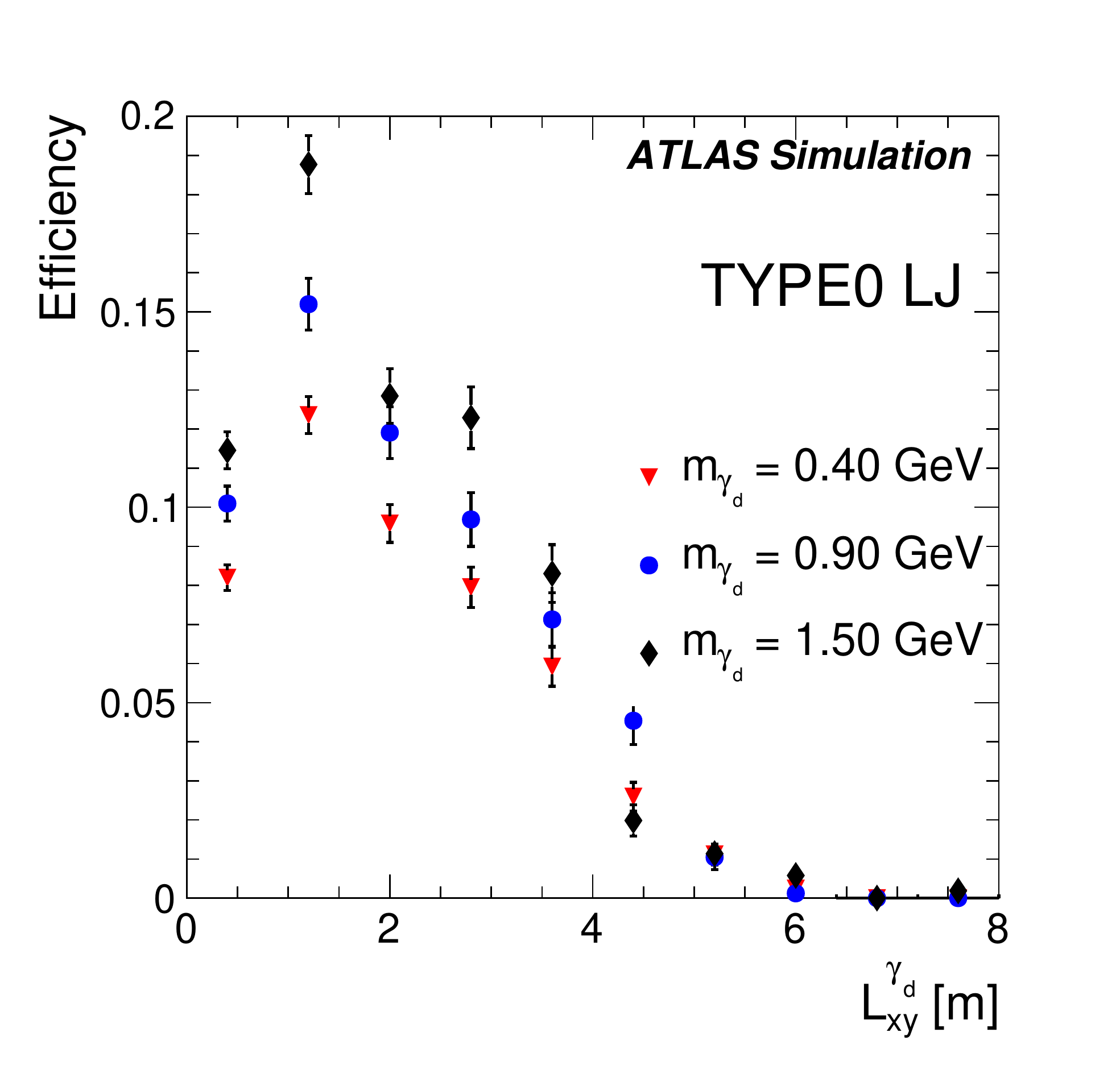}
\caption{Reconstruction efficiency of TYPE0 LJs as a function of $p_{\mathrm{T}}$ (left) and \lxy~(right) of the \gammad for \gdmumu~obtained from the LJ gun MC samples with \gammad masses 0.4, 0.9 and 1.5~\mygev. The uncertainties are statistical only.}
\label{fig:type0_recoeff_dp}
\end{figure}

\begin{figure}[t!]
\centering
\includegraphics[width=70mm]{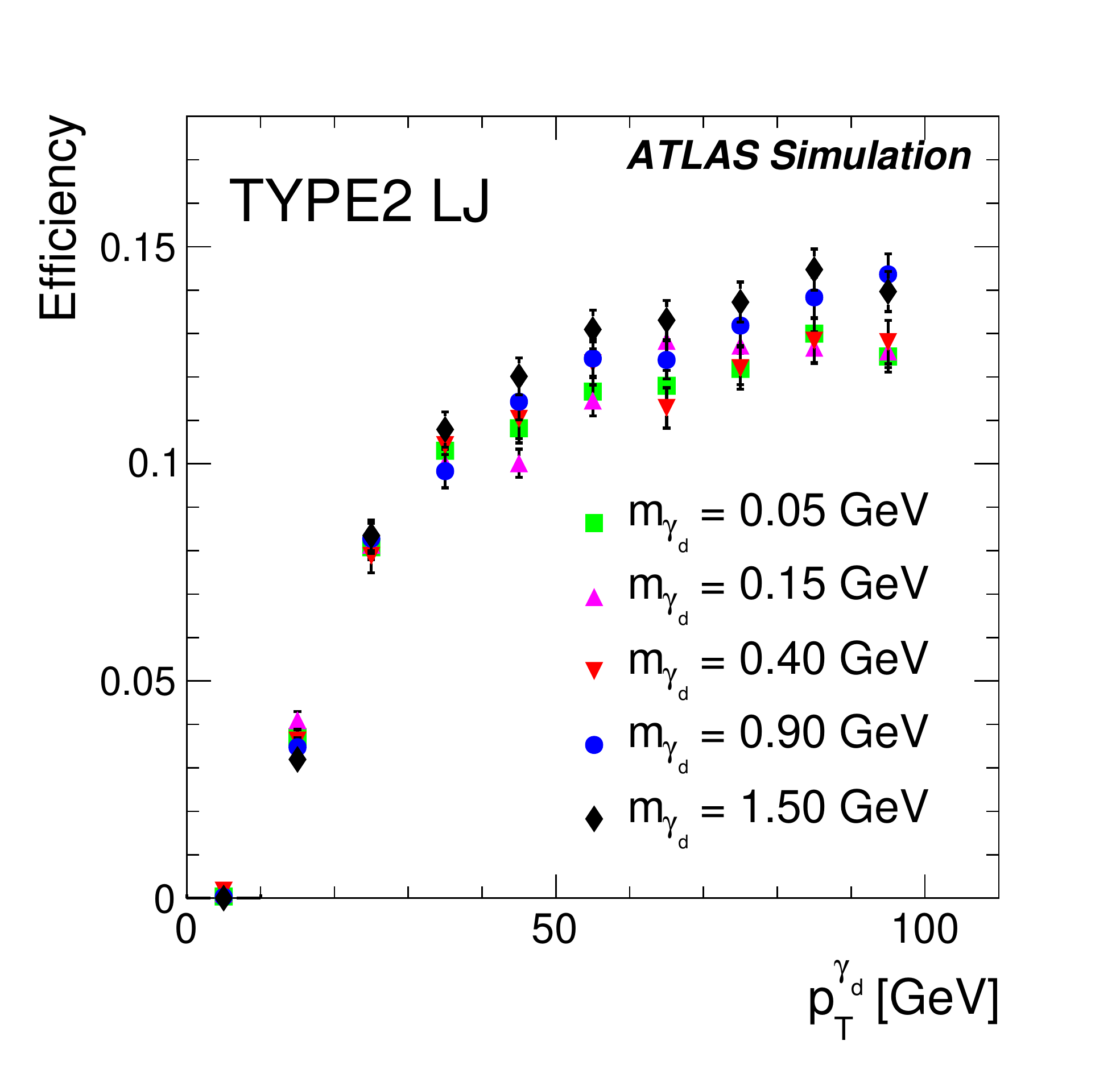}
\includegraphics[width=70mm]{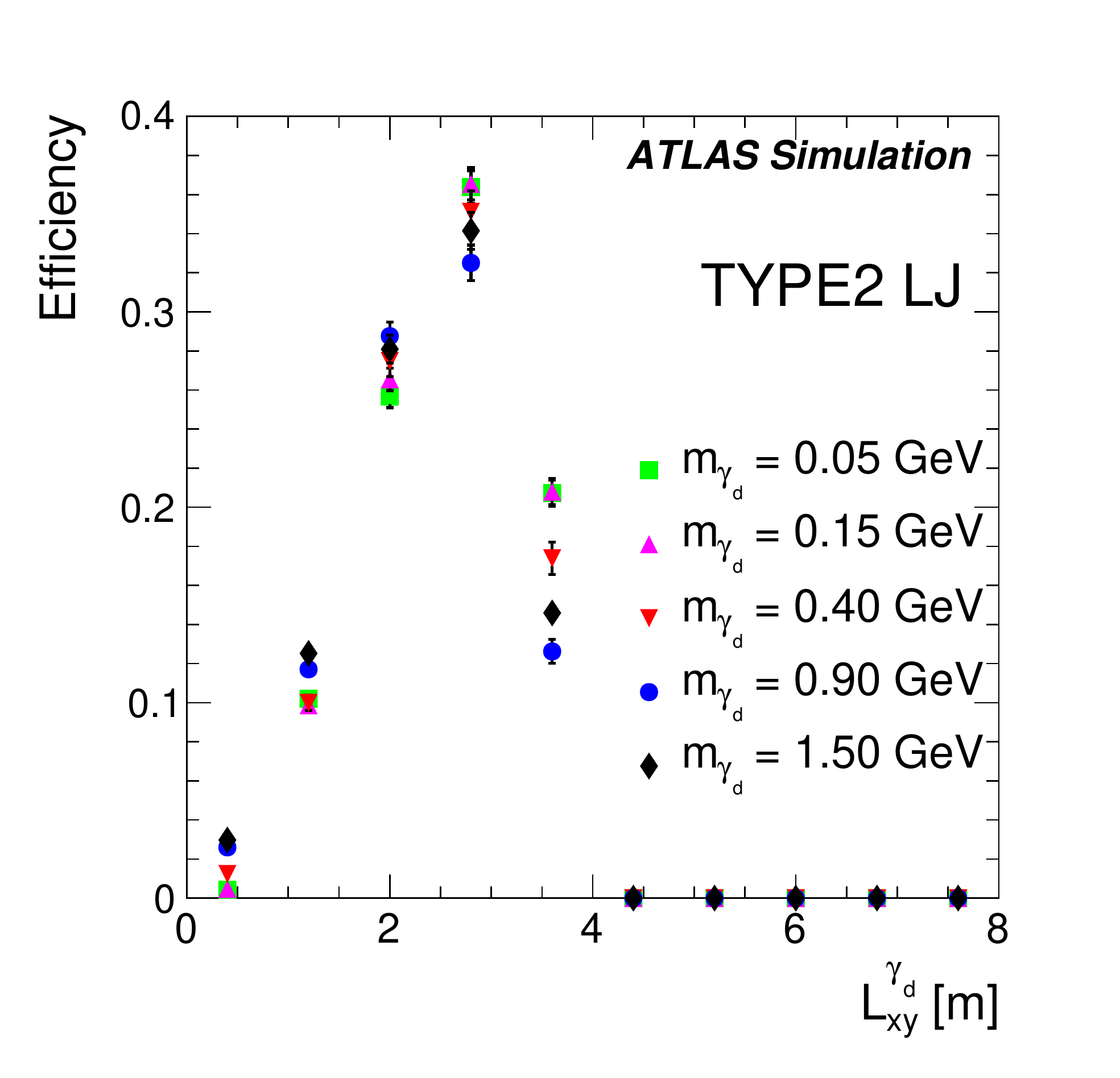}
\caption{Reconstruction efficiency of TYPE2 LJs as a function of $p_{\mathrm{T}}$ (left) and \lxy~(right) of the \gammad for \gdepai~obtained from the LJ gun MC samples with \gammad masses 0.05, 0.15, 0.4, 0.9 and 1.5~\mygev. The uncertainties are statistical only.}
\label{fig:type2_recoeff_dp}
\end{figure}

\begin{figure}[ht!]
\centering
\includegraphics[width=70mm]{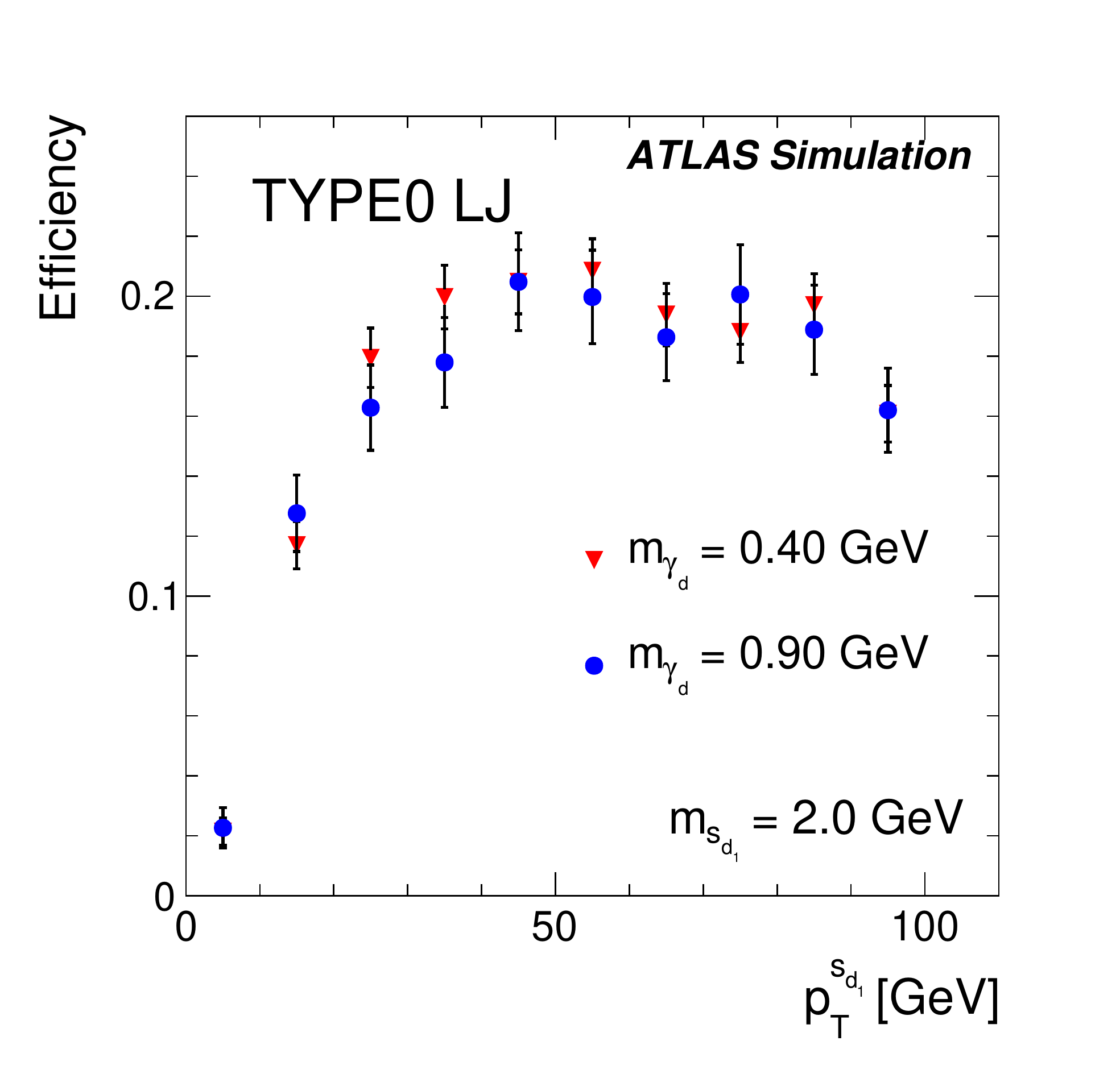}
\includegraphics[width=70mm]{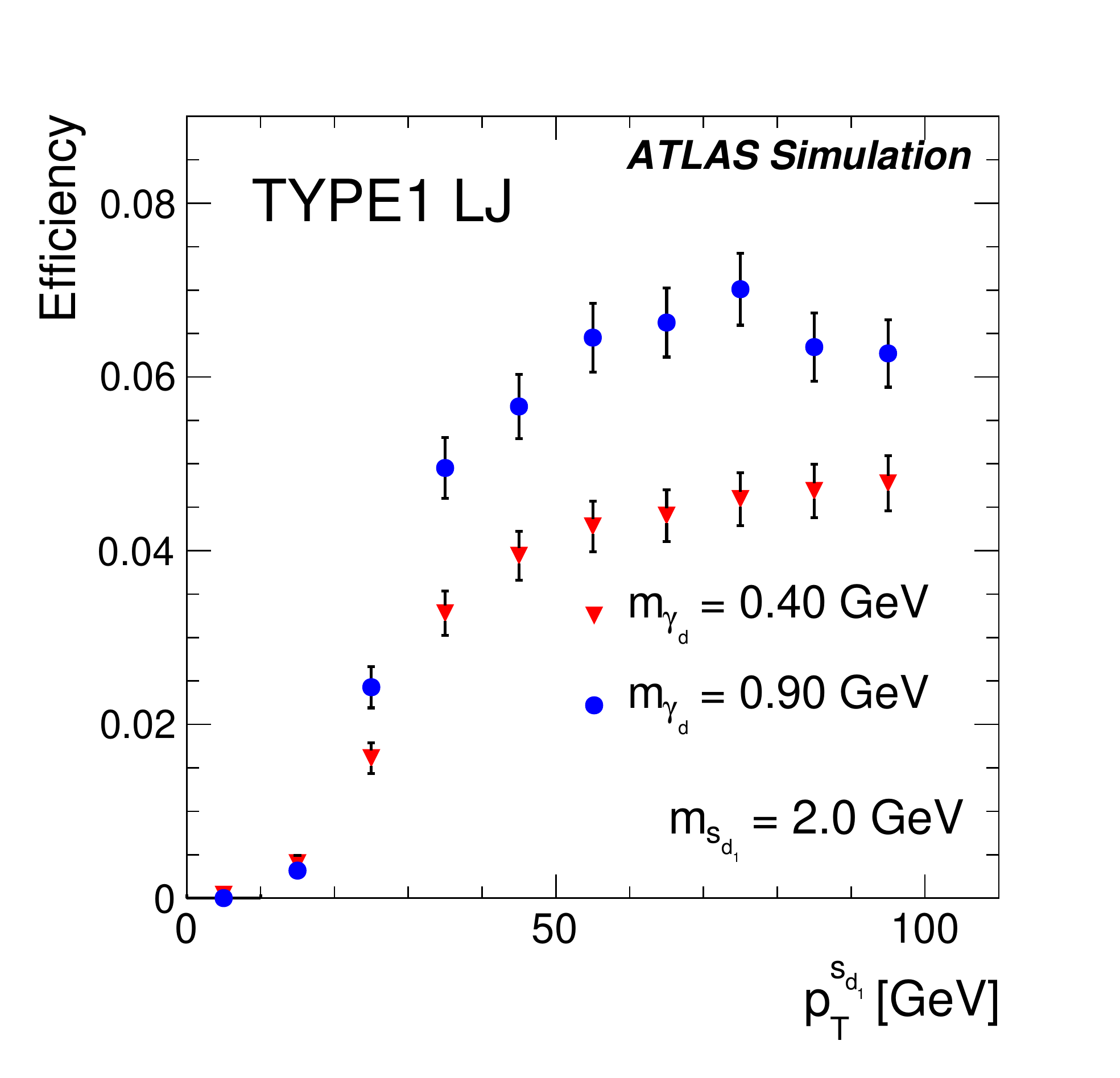}
\includegraphics[width=70mm]{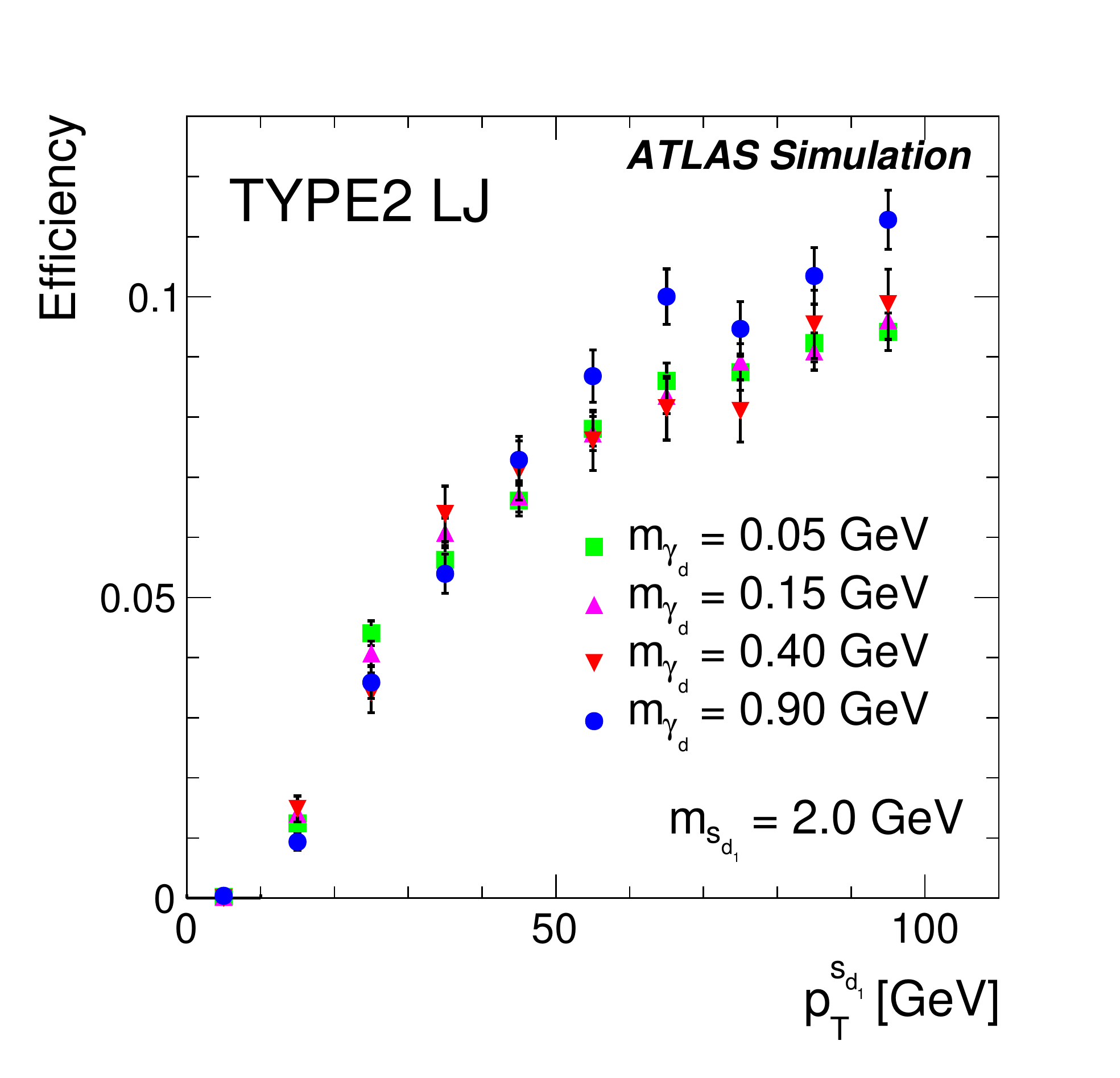}
\caption{Reconstruction efficiency of TYPE0 (top left), TYPE1 (top right) and TYPE2 (bottom) LJs as a function of the $p_{\mathrm{T}}$ of the \scalar for \LJs with two \gammads for an \scalar mass of 2~\mygev. For the \gammad, only the kinematically allowed masses are considered. The distributions for the other \scalar masses are very similar. The uncertainties are statistical only.}
\label{fig:recoeff_sd}
\end{figure}

In this section the \LJ reconstruction efficiency using the \LJ gun MC samples is presented. The reconstruction efficiency is given for \LJ with only one \gammad as a function of the \pt~and of the transverse decay distance \lxy~of the \gammad at the generation level. The efficiency is defined as the ratio of the number of reconstructed \LJs of a given type, without any trigger requirement, to the corresponding number of generated ones, of the same type, in a given \pt~or \lxy~interval. For \LJs with two \gammads, the reconstruction efficiency is presented as a function of the $p_{\mathrm{T}}$ of the \scalar. All the background rejection criteria defined in section~\ref{sec:LJbkg} are applied to the reconstructed \LJs. \\
Figure~\ref{fig:type0_recoeff_dp} shows the reconstruction efficiency for TYPE0 LJ as a function of the \pt~(left) and \lxy~(right) of the \gammad from \LJ gun MC samples with \gammad masses 0.4, 0.9 and 1.5~\mygev. \LJ gun MC samples with only one \gammad (\gdmumu) are used. As expected the efficiency decreases for $p_{\mathrm{T}}$~$\leq$~30~\mygev due to one of the two muons of the decay losing all its energy inside the calorimeters and decreases at high values of $p_{\mathrm{T}}$~due to the smaller opening angle between the two muons. The efficiency also decreases with increasing distance \lxy~from the primary vertex. This has two causes: the algorithm for reconstructing particle tracks in the MS has a loose requirement of extrapolation to the IP and the opening angle between the two muons decreases as the boost of the \gammad increases. The efficiency decrease at low \lxy~is due to the isolation requirement, which rejects the LJ if the muon tracks are reconstructed in the ID.
\\
Figure~\ref{fig:type2_recoeff_dp} shows the reconstruction efficiency for TYPE2 LJs as a function of the $p_{\mathrm{T}}$~(left) and \lxy~(right) of the \gammad from \LJ gun MC samples with \gammad masses 0.05, 0.15, 0.4, 0.9 and 1.5~\mygev. \LJ gun MC samples with only one \gammad (\gdepai) are used. As a consequence of the requirement on the EM fraction, mainly decays inside the HCAL are reconstructed.
Figure~\ref{fig:recoeff_sd} shows the reconstruction efficiency of TYPE0 LJs (top left), TYPE1 LJs (top right) and TYPE2 LJs (bottom) as a function of the $p_{\mathrm{T}}$ of the \scalar, obtained from the \LJ gun MC samples with an \scalar mass of 2~\mygev and kinematically allowed \gammad masses.
Only \LJ gun MC samples with two \gammads in the final state are used. The efficiency distributions are compatible with those obtained from the single \gammad samples.\footnote{ In case of two \gammads in the same \LJ, if one \gammad decays in electrons/pions before the HCAL, the \LJ is rejected due to the low EM fraction requirement.}
%
%######################## S U B - S E C T I O N ######################################
\subsection{LJ trigger efficiency}
\label{sec::LJtrig}
\begin{figure}[ht!]
\centering
\includegraphics[width=70mm]{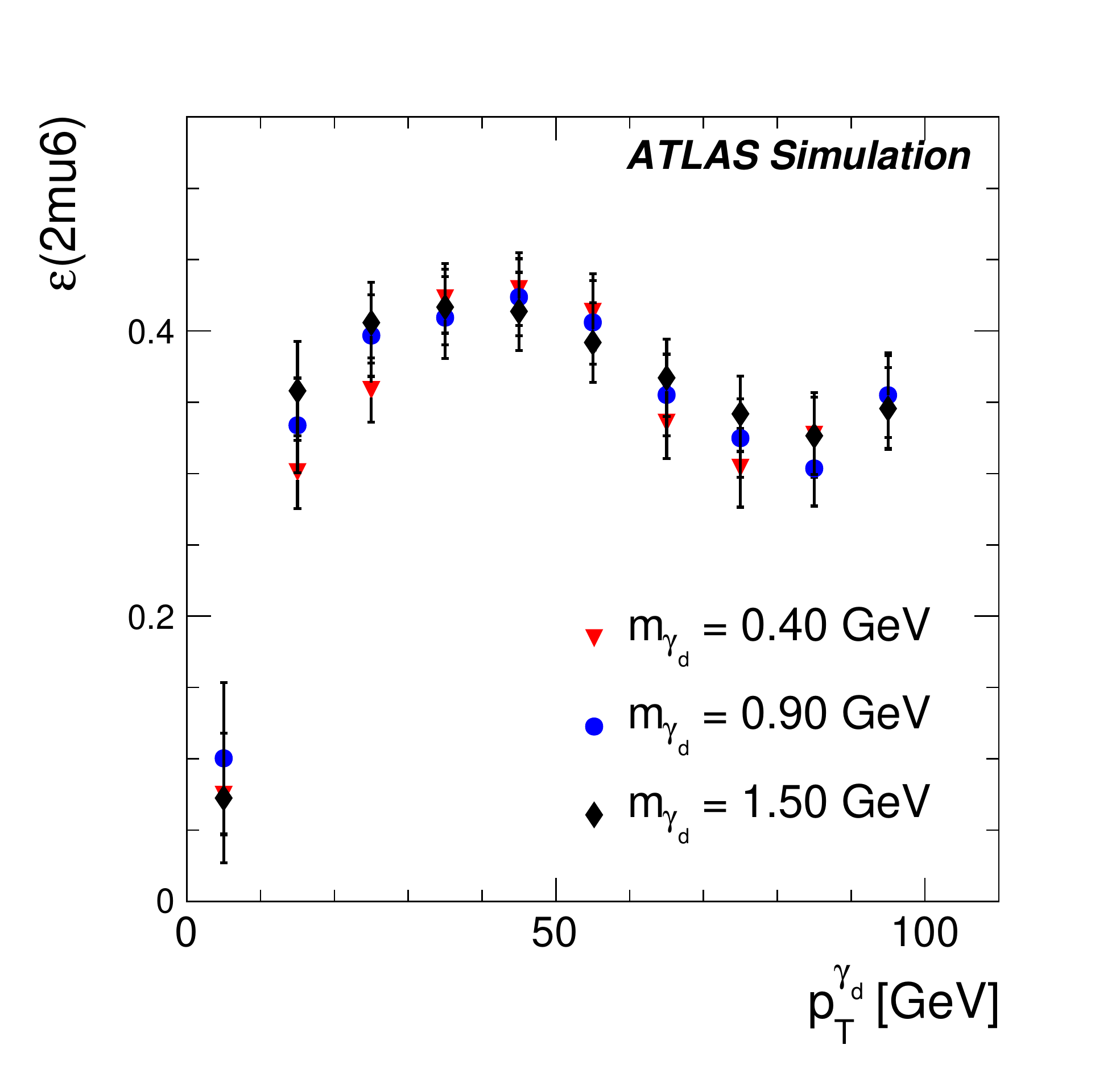}
\includegraphics[width=70mm]{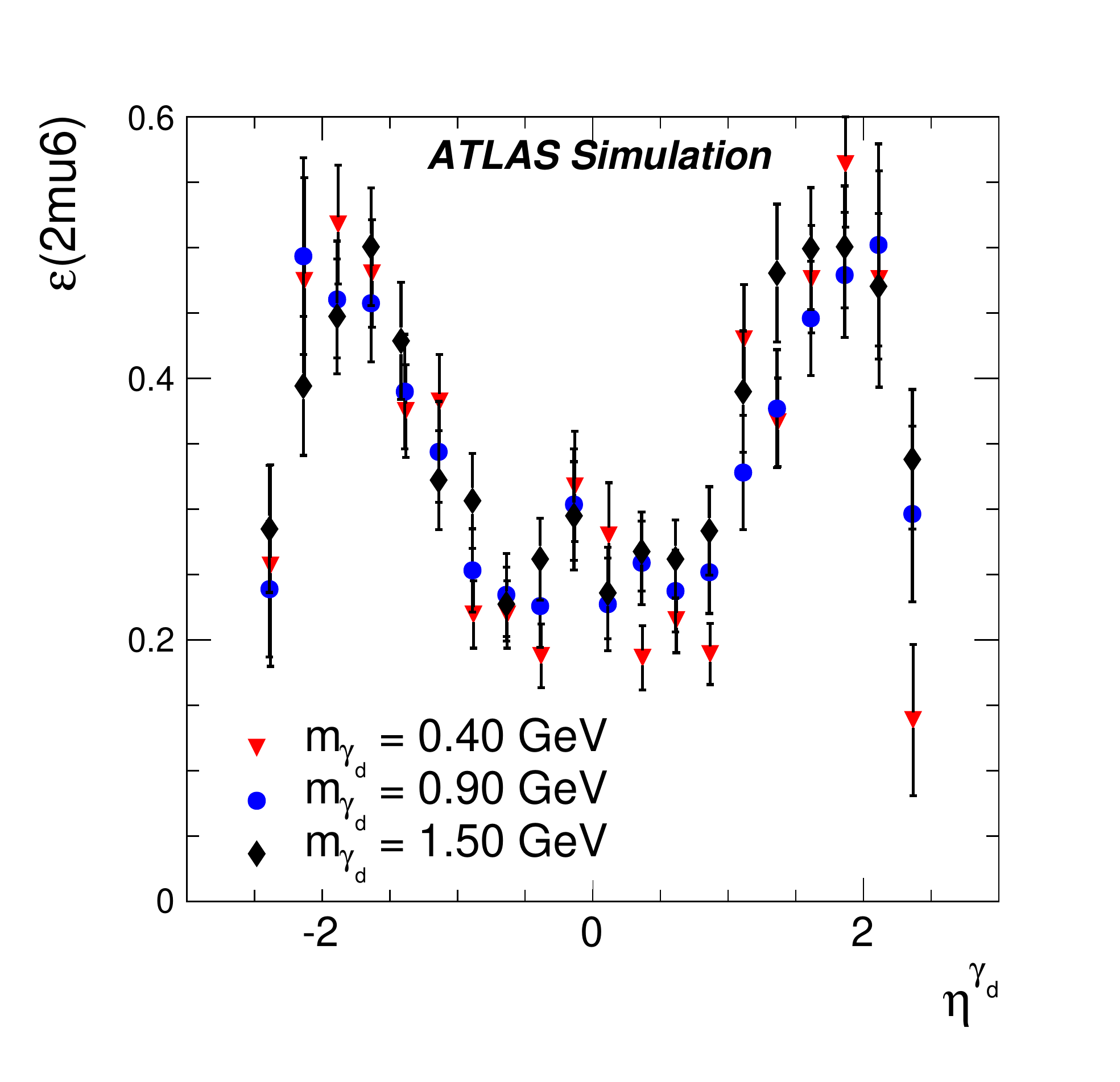}
\caption{Muon trigger efficiency, $\rm \varepsilon$(2mu6), as a function of \pt~(left) and $\eta$~(right) of the \gammad for \gdmumu~obtained from the LJ gun MC samples with \gammad masses 0.4, 0.9 and 1.5~\mygev. The uncertainties are statistical only.}
\label{fig:dimuon_trigeff_dp}
\end{figure}

\begin{figure}[ht!]
\centering
\includegraphics[width=70mm]{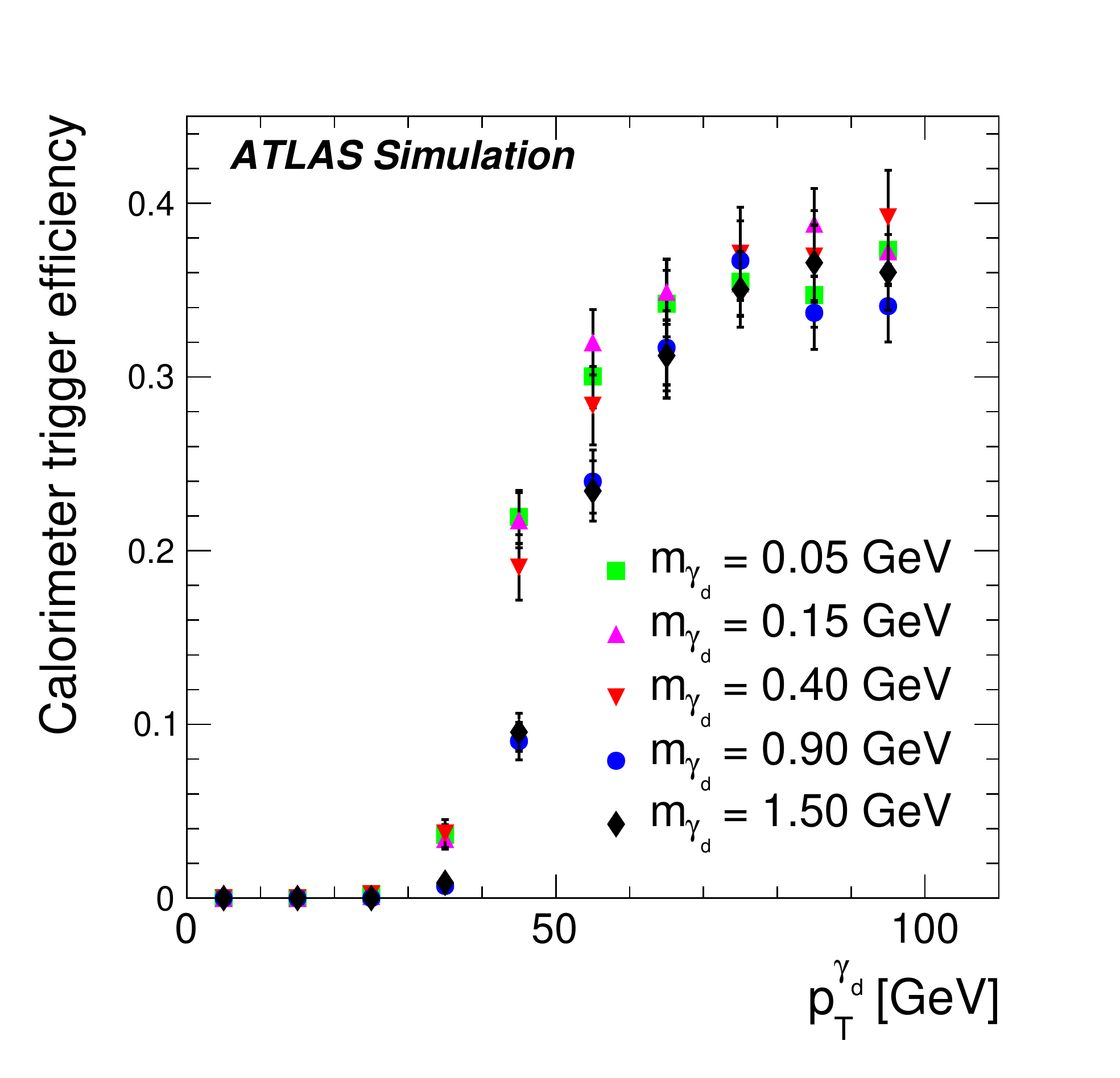}
\includegraphics[width=70mm]{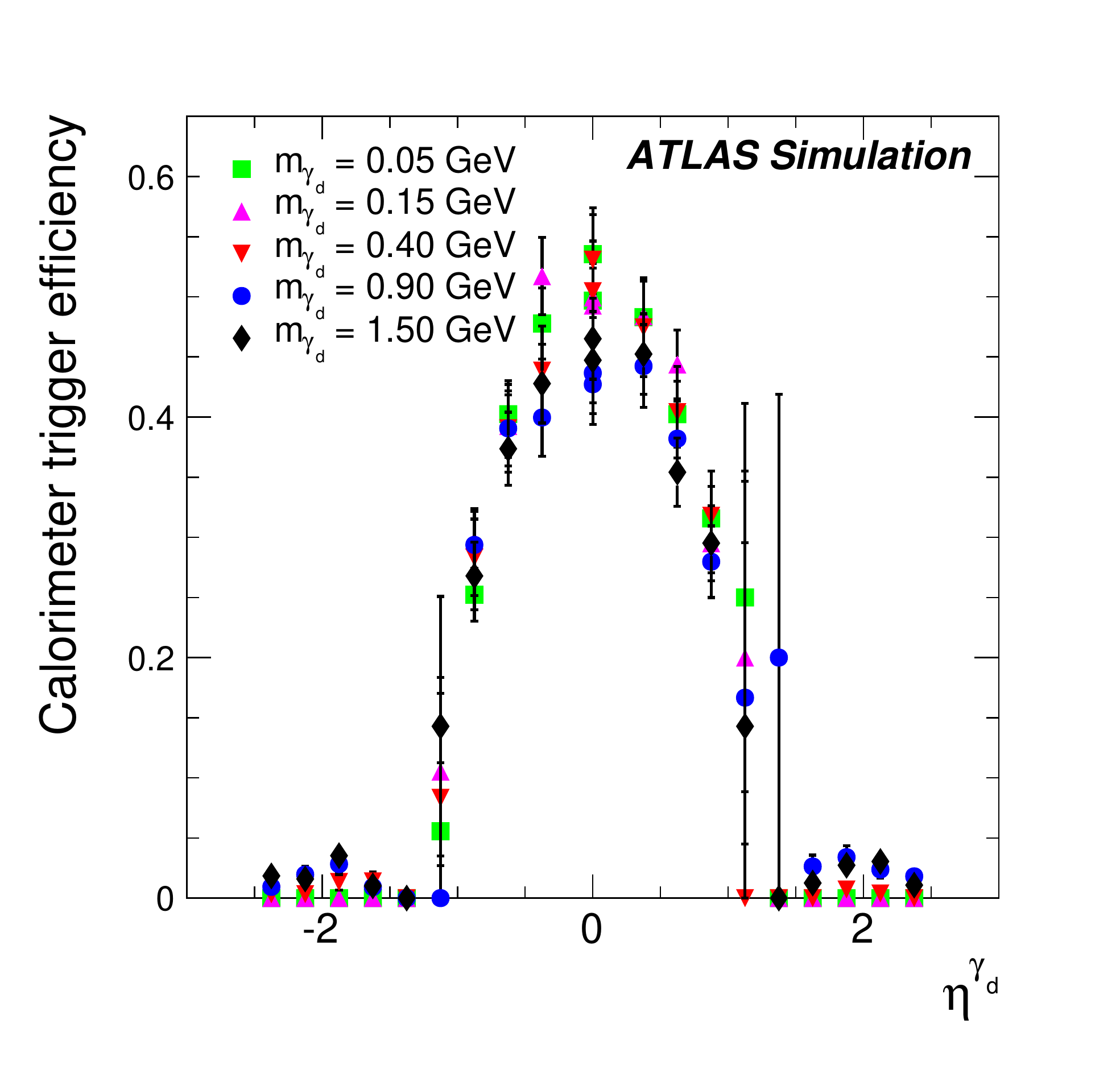}
\caption{Calorimetric trigger efficiency as a function of \pt~(left) and $\eta$ (right) of the \gammad for \gdepai~obtained from the LJ gun MC samples with \gammad masses 0.05, 0.15, 0.4, 0.9 and 1.5~\mygev. Similar distributions are obtained for \LJs containing two dark photons. The uncertainties are statistical only.}
\label{fig:type2_trigeff_dp}
\end{figure}

The trigger efficiency for events containing two displaced \LJs can be evaluated only at event level, i.e. taking into account the trigger response to both \LJs. However \LJ gun MC samples can provide information on the trigger efficiency for a single \gammad; from this efficiency the trigger behaviour for the full event can be easily derived.\\
A large fraction of the ATLAS muon triggers are strictly linked to the primary vertex and therefore are very inefficient in selecting tracks arising from displaced decay vertices. Selection of displaced \LJs of TYPE0 and TYPE1 needs an unprescaled multi-muon trigger that does not require matching between the muon track and an ID track and has a relatively low \pt~threshold. The only available HLT trigger in 2012 data taking satisfying these specifications requires at least three reconstructed muons in the MS with \pt~$\ge$~6~\mygev (3mu6 trigger). This multi-muon trigger requires, for an event containing two \gammads, one \gammad producing two \roi s and the other at least one. Therefore the efficiency of the trigger depends on the opening angle $\Delta R$ between the two muons from the \gammad decay. If the opening angle is smaller than the trigger granularity (see section~\ref{sec:ATLAS}), the L1 selects only one \roi. Therefore the probability for a single \gammad to produce two distinct \roi s is needed in order to evaluate the trigger efficiency. \\
Figure~\ref{fig:dimuon_trigeff_dp} shows the muon trigger efficiency, $\rm \varepsilon$(2mu6), for \gdecamu~obtained from the \LJ gun MC samples with \gammad masses 0.4, 0.9 and 1.5~\mygev, as a function of \pt~(left) and $\eta$ (right) of the \gammad. The efficiency $\rm \varepsilon$(2mu6) is defined as the fraction of \gdecamu~passing the offline selection that also satisfy the 2mu6 trigger. The decrease at high \pt~reflects the loss of trigger efficiency in the MS barrel when the boost of the \gammad increases: the angular separation between the muons decreases reducing the probability of two distinct \roi s. The effect of higher trigger granularity in the endcap relative to the barrel is clearly visible in figure~\ref{fig:dimuon_trigeff_dp}~(right).
\\
An estimate of the overall trigger efficiency per event, $\rm \varepsilon$(3mu6), can be derived from the $\rm \varepsilon$(2mu6) obtained with the \LJ gun MC samples. The probability of satisfying 3mu6 in events with two \gammads, is given by:
\begin{equation}
 \rm   p^{3mu6} = 2 \cdot \varepsilon(1mu6) \cdot \varepsilon(2mu6) - \varepsilon(2mu6) \cdot \varepsilon(2mu6)
  \label{eq:effi3mu6}
\end{equation}
where $\rm \varepsilon(1mu6)$ and $\rm \varepsilon(2mu6)$ are the probabilities for a \gammad to generate a 1mu6 and 2mu6 trigger, respectively. The $\rm \varepsilon(1mu6)$ can be assumed to be the single-muon trigger efficiency (80$\%$ in the barrel and 90$\%$ in the endcap part of the muon spectrometer). \\
In order to select displaced TYPE2 \LJs a single jet trigger with low EM fraction can be used~\cite{Trignote}.  The L1 trigger requires at least 40~\mygev energy deposition in a narrow region  \mbox{0.1$\times$0.1} (\mbox{$\Delta\eta\times\Delta\phi$}) of the calorimeters.  At L2 a cut $\le$ 0.06 on the EM fraction of the jet is applied. In addition, the trigger requirements for the jets are: \myet~>~30~\mygev, $|\eta|\le$~2.5 and no ID tracks with $\pt~>$~1.0~\mygev in the region 0.2~$\times$~0.2 ($\Delta\eta$~$\times$~$\Delta\phi$) around the jet axis. Finally, the EF requires the reconstructed jet to have  \myet~>~35~\mygev and applies beam-halo removal.\footnote{Uncalibrated calorimetric energy measurement is used in the three trigger levels.} \\
Figure~\ref{fig:type2_trigeff_dp} shows the calorimetric trigger efficiency for \gdepai~obtained from the \LJ gun MC samples with \gammad masses 0.05, 0.15,  0.4, 0.9 and 1.5~\mygev as a function of \pt~(left) and $\eta$ (right) of the \gammad. This efficiency is defined as the fraction of \gdepai~passing the offline selection that also satisfy the calorimetric trigger. The sharp decrease of the efficiency for $\pt~<$~60~\mygev is due to the L1 trigger requirement \myet~>~40~\mygev. The drop to zero for $|\eta|>1.0$ is due to the noisy-cell removal in the endcap hadronic calorimeter at trigger level~\cite{JetCleaning}.\footnote{A \gammad decay in the endcap HCAL is in general contained in a single cell. Most mis-reconstructed jets are caused by sporadic noise bursts in the endcap HCAL, where most of the energy is in single calorimeter cells, with often some cross-talk in neighbouring cells. Jets reconstructed from these problematic channels are considered fake jets and tagged as noise.}
%
%%%%%%%%%%%%%%%%%%%%%%%%% S E C T I O N %%%%%%%%%%%%%%%%%%%%%%%%%%%%%%%%%%%%%%%%%
\section{Event selection and backgrounds}
\label{sec:EvSel_and_Bkg}
%
%######################## S U B - S E C T I O N ######################################
\subsection{Data and background samples}
\label{sec:Data_bkg_samples}
The data used for this analysis were collected during the entire 2012 data-taking period and selected by the logical OR of the two triggers described in section~\ref{sec::LJtrig}. Only data in which all the ATLAS subdetectors were running at nominal conditions were selected. The total integrated luminosity corresponds to \intlumi. \\
Potential backgrounds include all processes that lead to prompt muons with or without associated jets such as the SM processes {\it W}+jets, {\it Z}+jets, \ttbar, single-top, {\it WW}, {\it WZ}, and {\it ZZ}. The MC samples used to estimate the prompt lepton background are generated using \pythia~\cite{PYTHIA8}~({\it W}+jets and {\it Z}+jets) and \mcatnlo~\cite{mcatlno}~(\ttbar, {\it WW}, {\it WZ}, and {\it ZZ}). The generated MC events are processed through the full ATLAS simulation and reconstruction chain. Additional {\it pp} interactions in the same and nearby bunch crossings (pile-up) are included in the simulation. All MC samples are re-weighted to reproduce the observed distribution of the number of interactions per bunch crossing in the data. \\
Cosmic rays in ATLAS come mostly from the skyward direction and arrive mainly from the two large access shafts to the pit. Cosmic-ray muons interact with the detector as minimum-ionizing particles and most traverse the entire detector. In some cases, cosmic rays can produce large energy deposits in the calorimeter system. These may be reconstructed as jets, which result in a background to the TYPE1 and TYPE2 \LJ selections used in this analysis. Moreover, muon bundles in cosmic-ray air showers can mimic the signature of TYPE0 \LJs.\footnote{Muon bundles are showers of high-multiplicity quasi-parallel penetrating particles produced by very high-energy cosmic rays.} The same triggers used to select the data sample in the collisions were also active in the 2012 data taking in the empty bunch crossings. Such data are used to study and to estimate the cosmic-ray background to the signal.
%
%######################## S U B - S E C T I O N ######################################
\subsection{Selection of events with \LJs}
The selection of events starts by requiring at least two reconstructed \LJs (see section \ref{sec:LJdef}). The requirements for the individual \LJ background rejection (see section~\ref{sec:LJbkg}) are then applied to the selected events.
At the event level, additional requirements are made to separate the \LJ signal from background.
%
% -:-:-:-:-:-:-:-:-:-:-:-:-:-:-:-:-:-:-:P A R A G R A P H :-:-:-:-:-:-:-:-:-:-:-:-:-:-:-:-:-:-:-:-:-:-:-:-:-:-
\paragraph{LJ isolation}
All the \noprompt~\LJs have to be isolated in the ID. As a global variable for the \LJ event selection, the highest ID \SumpT~(see section~\ref{sec:LJbkg}) of the \LJs in the event (denoted by \maxsumpt~in the following) is required to be $\leq 3$~\mygev.
%
% -:-:-:-:-:-:-:-:-:-:-:-:-:-:-:-:-:-:-:P A R A G R A P H :-:-:-:-:-:-:-:-:-:-:-:-:-:-:-:-:-:-:-:-:-:-:-:-:-:-
\paragraph{LJ production}
In order to reduce the background level in the \LJ event selection, an additional requirement on the azimuthal angle $\mydelphi$~between the two \LJs is introduced. A \myabsdelphi~$\geq 1$ requirement significantly reduces the background without large signal losses even in models where \LJ production is not back-to-back~\cite{alphad}.
\\
\\
The complete list of the criteria for the selection of events with \LJs is summarized in table~\ref{tab:cutFlow} and the number of events observed in data applying the \LJ selection is shown in table~\ref{tab:cutFlowdata}.
\begin{table*}[t!]
{
\small
\resizebox{\columnwidth}{!}{
\centering
\begin{tabular}{ | c | c | }
\hline
 Requirement&  Description\\
\hline
\hline
 Two reconstructed \LJs            & select events with at least two reconstructed \LJs                  \\
\hline
  $\eta$ range (TYPE1)              & remove jets with $|\eta|>2.5$                                              \\
\hline
  $\eta$ range (TYPE2)              & remove jets with $|\eta|>2.5$  and $1.0<|\eta|<1.4$          \\
\hline
EM fraction  (TYPE2)                  & require EM fraction of the jet $<0.1$                                    \\
\hline
Jet width W  (TYPE2)                  & require width of the jet $<0.1$                                             \\
\hline
 Jet timing (TYPE1/TYPE2)         & require jets with timing $-$1~ns~$<t<$~5~ns                     \\
\hline
 NC muons (TYPE0/TYPE1)        & require muons without ID track match                                   \\
\hline
 ID isolation                               & require \maxsumpt~$\leq 3$~\mygev                                  \\
\hline
 $\Delta\phi$                            & require \myabsdelphi~$\geq 1$~rad between the two \LJs    \\
\hline
\end{tabular}
}
}
\caption{Requirements for selection of events with \LJs. The requirements are applied to all LJ types unless otherwise specified.}
\label{tab:cutFlow}
\end{table*}
\begin{table}[t!]
\centering
\small
\begin{tabular}{ | c | c | c | c | c | c | c | c |}
\hline
                LJ pair types              &   0-0   &   0-1  &    0-2   &   1-1   &   1-2   &    2-2  &  All                           \\
\hline
\hline
Trigger selection                       &   \multicolumn{7}{c|}  {9.226$\times10^{6}$}                                                \\
\hline
Good primary vertex                 &     \multicolumn{7}{c|}  { 9.212$\times10^{6}$}                                             \\
\hline
Two reconstructed \LJs             &   946   &  1771 & 16676 &   1382 & 19629 &  82653& 123057                     \\
\hline
$\eta$ range (TYPE1/TYPE2)    &    946  &  1269 &   5063 &   701   &  3838  & 25885 & 37702                        \\
\hline
EM fraction  (TYPE2)                 &    946  & 1269  &   393   &   701   &    172  &  4713  &  8194                         \\
\hline
Jet width W  (TYPE2)                 &   946  & 1269  &   350    &    701  &   148   &  3740 &  7154                          \\
\hline
 Jet timing (TYPE1/TYPE2)        &    946  & 1054 &    216   &    547  &     92   &    578  &  3433                          \\
\hline
NC muons  (TYPE0/TYPE1)       &      27  &    3    &     42    &       5   &     5     &   578  &  660                             \\
\hline
ID isolation                              &      12  &     0    &    19     &     4     &     3     &  160   &  198                            \\
\hline
\myabsdelphi                          &      11  &     0    &     11    &     4     &     3     &    90    &  119                            \\
\hline
\end{tabular}
\caption{Number of selected data events at different stages of the selection process and for each of the LJ pair types, for the full 2012 data sample.}
\label{tab:cutFlowdata}
\end{table}
%
%######################## S U B - S E C T I O N ######################################
\subsection{Background evaluation}
\label{sec:ABCDQCD}
%
% -:-:-:-:-:-:-:-:-:-:-:-:-:-:-:-:-:-:-:P A R A G R A P H :-:-:-:-:-:-:-:-:-:-:-:-:-:-:-:-:-:-:-:-:-:-:-:-:-:-
\paragraph{Cosmic-ray background}
The nominal LHC configuration for proton--proton collisions contains 3564 bunch crossings per revolution. Not all bunches are actually filled with protons. Empty bunch crossings contain no protons and allow for the study of cosmic-ray background events. The \LJ selection for events triggered in the empty bunch crossings, using the same triggers as the ones used to select the data,  is shown in table~\ref{tab:cutFlowEMPTY}. The selection criteria used are identical to the ones employed for the filled bunch crossings, except for applying a primary vertex requirement. The ratio of filled to empty bunch crossings is used to rescale the observed number of events to the {\it pp} collision data.
\begin{table}[t!]
\centering
\small
\begin{tabular}{| c | c | c | c | c | c | c | c |}
\hline
              \LJ pair types &   0-0   &   0-1  &    0-2   &   1-1   &   1-2   &    2-2  &  All                             \\
\hline
\hline
Trigger selection                         &   \multicolumn{7}{c|}  {161951 }                                                                         \\
\hline
Good primary vertex                   &     \multicolumn{7}{c|}  { not applicable }                                                             \\
\hline
 Two reconstructed \LJs                &        6     &       0    &      42   &       0     &    36      &   3744   &  3838               \\
\hline
 $\eta$ range (TYPE1/TYPE2)     &        6     &       0     &     29   &       0     &     17     &    2243   &  2295                 \\
\hline
 EM fraction  (TYPE2)                  &        6     &       0     &     29    &        0     &    17     &    2190  &  2242                 \\
\hline
 Jet width W  (TYPE2)                  &       6     &        0    &      22    &       0     &       6    &      1632 &  1666                 \\
\hline
 Jet timing (TYPE1/TYPE2)          &        6     &        0    &       6    &       0     &       0     &      24    &  36                      \\
\hline
 NC muons (TYPE0/TYPE1)         &        6      &        0    &       6    &       0     &      0     &       24    &  36                     \\
\hline
 ID isolation                               &        6      &        0    &       6    &       0     &      0     &        24   & 36                      \\
\hline
 \myabsdelphi                           &        6      &         0   &       5     &       0     &      0     &         4   & 15                      \\
\hline
 Rescaled to interactions            & $15\pm6$ & $0^{+3.1}_{-0}$ &  $14\pm6$ & $0^{+3.1}_{-0}$ &  $0^{+3.1}_{-0}$ &         $11\pm7$ & $40\pm10$           \\
\hline
\end{tabular}
\caption{Result of applying the \LJ selection to events triggered in the empty bunch crossings. Number of selected data events at different stages of the selection process and for each \LJ pair types. Except for the last row, all these numbers are not rescaled by the ratio of filled to empty bunches in the LHC operation. The quoted uncertainties are statistical only.}
\label{tab:cutFlowEMPTY}
\end{table}
After rescaling, the estimated background contribution to the full 2012 dataset is $40\pm10$, as shown in the last row of table~\ref{tab:cutFlowEMPTY} where the quoted uncertainties are statistical only.
%
% -:-:-:-:-:-:-:-:-:-:-:-:-:-:-:-:-:-:-:P A R A G R A P H :-:-:-:-:-:-:-:-:-:-:-:-:-:-:-:-:-:-:-:-:-:-:-:-:-:-
\paragraph{Background from electroweak and $t\bar t$ processes}
All these MC background samples give negligible contributions even at the trigger level.

\begin{figure}[t!]
\centering
\includegraphics[width=95mm]{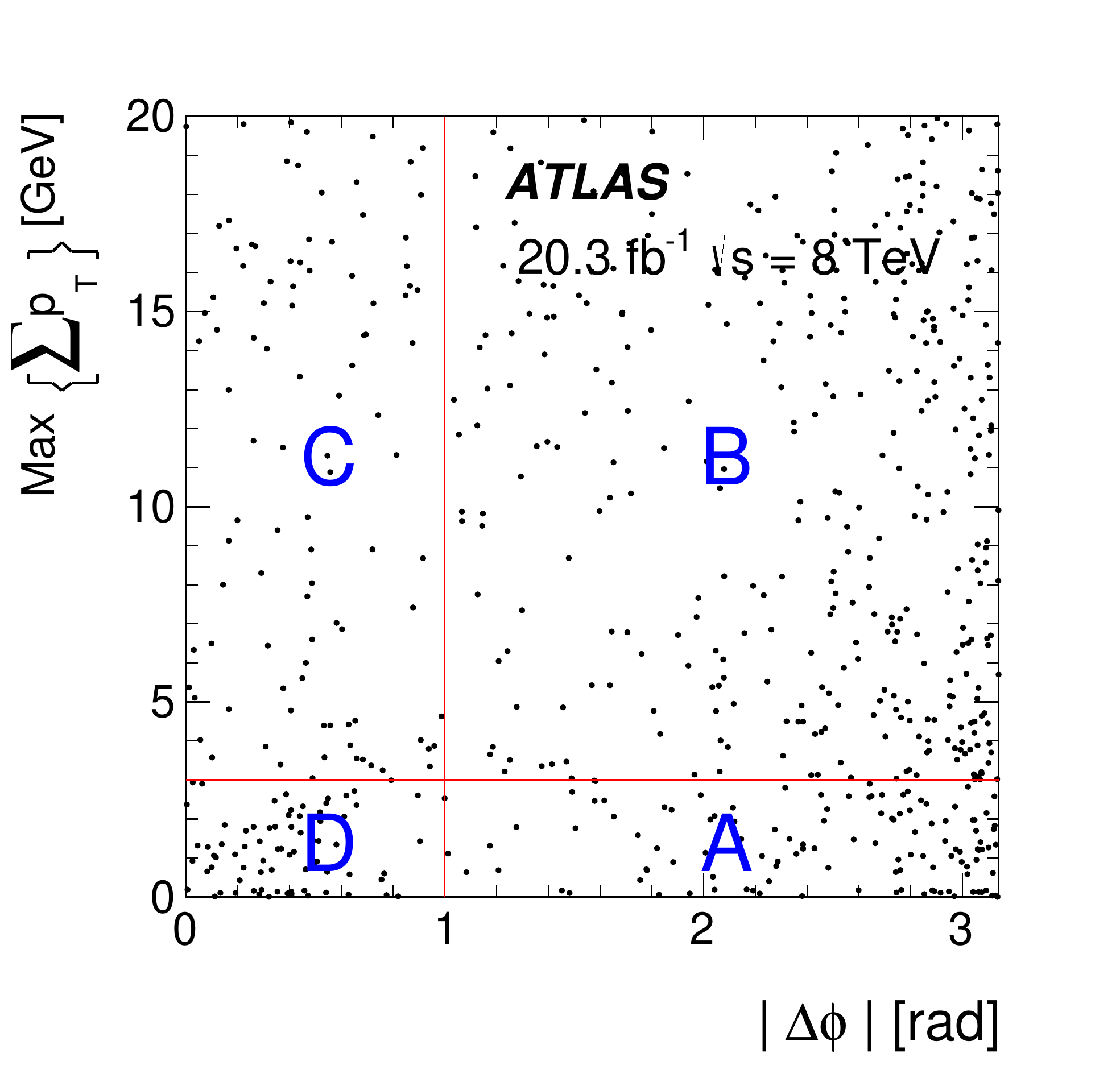}
\caption{Distribution of \LJ events in the ABCD plane before the requirements \maxsumpt~$\leq$~3~\mygev and $|\Delta\phi|~\geq$~1.}
\label{fig:ABCD}
\end{figure}

%
% -:-:-:-:-:-:-:-:-:-:-:-:-:-:-:-:-:-:-:P A R A G R A P H :-:-:-:-:-:-:-:-:-:-:-:-:-:-:-:-:-:-:-:-:-:-:-:-:-:-
\paragraph{Multi-jet background using the ABCD method}
The multi-jet background evaluation is done using a data-driven (ABCD) method. This is a simplified matrix method that relies on the assumption that two relatively uncorrelated variables can be identified for the separation of signal from background. It is assumed that the multi-jet background distribution can be factorized in the \myabsdelphi, \maxsumpt~plane.  Figure~\ref{fig:ABCD} shows the event distribution in this plane before the requirements on \myabsdelphi~and \maxsumpt. If A is the signal region (\maxsumpt $\leq$~3~\mygev and \myabsdelphi~$\geq $ 1), the number of background events in A can be predicted from the population of the other three regions: $N_{\rm{A}} = N_{\rm{D}} \times N_{\rm{B}}/ N_{\rm{C}}$, assuming a negligible leakage of signal into regions B, C and D.
Table~\ref{tab:abcd} summarizes the observed yields in the data for the three regions B, C and D.
The cosmic-ray estimated values (using the cosmic-ray data collected in the empty bunches, rescaled for the filled-to-empty bunch ratio) in the same regions are given in the table; in this case the events in A are the expected ones from the cosmic-ray data, after rescaling. In order to evaluate the multi-jet background, the cosmic-ray contribution in region D is subtracted (cosmic rays are usually isolated); the estimated number of multi-jet background events in the signal region is $N_{\rm{A}} = 70 \pm 58 (stat)$.
\begin{table*}[t!]{
\small
\resizebox{\columnwidth}{!}{
\centering
\begin{tabular}{| c | c | c | c | c |}
\hline
Data Type  & Events in B & Events in C & Events in D & Expected Events in A \\
\hline
\hline
 Cosmic-ray data  & 0 & 0 &60 $\pm$ 13& 40 $\pm$ 10 \\
\hline
 Data (cosmic rays subtracted) & 362 $\pm$ 19& 99 $\pm$ 10& 19 $\pm$ 16& 70 $\pm$ 58\\
\hline
\end{tabular}
}}
\caption{Event yields in the four ABCD regions used to estimate the multi-jet background with the ABCD method in the \LJ signal region. All LJ pair types are used. The quoted uncertainties are statistical only.}
\label{tab:abcd}
\end{table*}
The expected multi-jet background in the signal region is strongly reduced by removing TYPE2-TYPE2 \LJ pairs from the selection. Without this LJ pair type, 29 events are observed in the signal region, corresponding to 24$\%$ of the total. The result of the background estimation obtained when removing TYPE2-TYPE2 is shown in table~\ref{tab:abcd-no22}.
\begin{table*}[tt!]{
\resizebox{\columnwidth}{!}{
\centering
\small
\begin{tabular}{| c | c | c | c | c |}
\hline
Data Type  & Events in B & Events in C & Events in D & Expected events in A \\
\hline
\hline
Cosmic-ray data  & 0 & 0 &3 $\pm$ 3& 29 $\pm$ 9 \\
\hline
Data (cosmic rays subtracted) & 29 $\pm$ 5& 15 $\pm$ 4& 6 $\pm$ 4& 12 $\pm$ 9\\
\hline
\end{tabular}
}}
\caption{Event yields in the four regions used to estimate the multi-jet background with the ABCD method in the \LJ signal region. TYPE2-TYPE2 \LJs are excluded. The quoted uncertainties are statistical only.}
\label{tab:abcd-no22}
\end{table*}
%
%%%%%%%%%%%%%%%%%%%%%%%%% S E C T I O N %%%%%%%%%%%%%%%%%%%%%%%%%%%%%%%%%%%%%%%%%
\section{Results for the FRVZ models}
\label{sec:LJresult}
In this section the data are interpreted in the context of the two FVRZ models as examples for the production of LJs.
%
%######################## S U B - S E C T I O N ######################################
\subsection{MC simulation of the FRVZ models}
\label{sec:MC-FRVZ}
The set of parameters used to generate the signal MC samples is listed in table~\ref{tab:param}.
The Higgs boson is generated through the gluon fusion production mechanism, which is the dominant production mechanism for a low-mass Higgs boson. The gluon fusion Higgs boson production cross section in {\it pp} collisions at \myrts = 8~\tev, estimated at next-to-next-to-leading order (NNLO)~\cite{HiggsCrossS}, is $\sigma_{\textrm\SMs} = $ 19.2 pb for $m_{H}=$~125~\mygev. The masses of $f_{\rm d_{2}}$ and HLSP are chosen to be light relative to the Higgs boson mass, and far from the kinematic threshold at $m_{\rm{HLSP}} + m_{\gamma_{\rm d}} = m_{f_{\rm d_{2}}}$.\footnote{No hidden-sector radiation is included in the generated samples, which corresponds to the choice $\alpha_d\lesssim 0.01$. This may affect the trigger and reconstruction efficiencies.} \\
For a dark photon mass of 0.4~\mygev, the \gammad decay branching ratios (BR) are expected to be 45$\%~\ee$, 45$\%~\mu^+\mu^-$ and 10$\%~\pi^+\pi^-$~\cite{b10b}. The mean lifetime $\tau$ of the \gammad (expressed as $\tau$ times the speed of light $c$) is a free parameter of the model. In the generated samples \ctau~=~47~mm  is chosen so that about 85$\%$ of the decays occur inside the trigger-sensitive ATLAS detector volume, i.e. up to 7~m in radius and $\pm$13~m along the $z$-axis. The detection efficiency is estimated for a range of \gammad mean lifetimes through re-weighting of the generated samples. \\
The \pythia~generator is used, linked together with \madgraph 5~\cite{b12} and \bridge~\cite{BRIDGE}, for gluon fusion production of the Higgs boson and the subsequent decay to hidden-sector particles. The generated MC events are processed through the full ATLAS simulation chain based on \geant~and then reconstructed.
\begin{table*}[ht!]{
\resizebox{\columnwidth}{!}{
\small
\centering
\begin{tabular}{| c | c | c | c | c | c | c | c | c | c | c |}
\hline
Model & Events & $m_{\rm h}$ & $m_{\rm f_{d_{2}}}$ & $m_{\rm HLSP}$ & $m_{\rm s_{d_{1}}}$  & $m_{\gamma_{\rm d}}$ & $c\tau_{\gamma_{\rm d}}$ & BR & BR & BR \\
  & & $[\rm \mygev]$  &$[\rm\mygev]$ &$[\rm\mygev]$ & $[\rm\mygev]$ & $[\rm\mygev]$ & [mm] & \gammad $\to ee$ & \gammad $\to \mu\mu$ & \gammad $\to \pi\pi$ \\
\hline
\hline
Two \gammads & 150k & 125& 5.0 & 2.0 & - & 0.4 & 47 & 0.45 & 0.45 & 0.10    \\
\hline
Four \gammads & 150k & 125& 5.0 & 2.0 & 2.0 & 0.4 & 47 & 0.45 & 0.45 & 0.10 \\
\hline
\end{tabular}
}}
\caption{Parameters of the FRVZ models used to generate the signal MC samples.}
\label{tab:param}
\end{table*}
%
%######################## S U B - S E C T I O N ######################################
\subsection{\LJ selection applied to FRVZ models}
Assuming a 10$\%$ BR of the Higgs boson to the hidden sector and a total integrated luminosity of \intlumi, the expected number of \LJ events for the two benchmark models are shown in table~\ref{tab:cutFlowFRVZ2} and table~\ref{tab:cutFlowFRVZ4}. Signals of 60 $\pm$ 7 (stat) and 104 $\pm$ 9 (stat) events are expected for the two-\gammad and four-\gammad FRVZ models, respectively.
\begin{table}[ht!]
\centering
\small
\begin{tabular}{| c | c | c | c | c | c | c | c |}
\hline
                LJ pair types &   0-0   &   0-1  &    0-2   &   1-1   &   1-2   &    2-2  &  All                                  \\
\hline
\hline
Total number of events                       &   \multicolumn{7}{c|}  { 39730 $\pm$ 100 } \\
\hline
Trigger selection                       &   \multicolumn{7}{c|}  { 1330 $\pm$ 30 } \\
\hline
Good primary vertex                 &     \multicolumn{7}{c|}  { 1330 $\pm$ 30 } \\
\hline
Two reconstructed \LJs                &   86     &     9     &   40     &     0    &     1    &   39     & 175 $\pm$ 7       \\
\hline
$\eta$ range (TYPE1/TYPE2)   &   86     &     8     &   27     &     0    &     1    &   23     & 145 $\pm$ 6        \\
\hline
EM fraction  (TYPE2)                 &   86     &     8     &   23     &     0    &     1    &   12     &  130 $\pm$ 6     \\
\hline
Jet width W  (TYPE2)                 &   86     &     8     &   23     &     0    &     1    &   12     &  130 $\pm$ 6      \\
\hline
Jet timing (TYPE1/TYPE2)        &    86     &     6     &   23     &     0    &     1    &   11     &  128 $\pm$ 6     \\
\hline
NC muons (TYPE0/TYPE1)        &    50     &     4     &   17     &     0    &     0    &   11     &    82 $\pm$ 5      \\
\hline
ID isolation                              &     37     &     2     &   13     &     0    &     0    &   10     &    63 $\pm$ 4       \\
\hline
\myabsdelphi                          & 35 $\pm$ 3 & 2 $\pm$ 1 & 12 $\pm$ 2 &  $0^{+0.6}_{-0}$  & $0^{+0.6}_{-0}$ & 10 $\pm$ 2 & 60 $\pm$ 4 \\
\hline
\end{tabular}
\caption{Expected number of \LJ events for the two-\gammad FRVZ model, using the parameter values in table~\ref{tab:param}. The numbers refer to selected signal events at different stages of the selection process and for each \LJ pair type. The number of signal events is rescaled to the \intlumi total integrated luminosity and the quoted uncertainties are statistical only. The detection efficiency is 1.5 $\times 10^{-3}$.}
\label{tab:cutFlowFRVZ2}
\end{table}
\begin{table}[ht!]
\centering
\small
\begin{tabular}{| c | c | c | c | c | c | c | c |}
\hline
           \LJ pair types               &   0-0   &   0-1  &    0-2   &   1-1   &   1-2   &    2-2  &  All                      \\
\hline
\hline
Total number of events                       &   \multicolumn{7}{c|}  { 39730 $\pm$ 100 } \\
\hline
Trigger selection                    &   \multicolumn{7}{c|}  { 2518 $\pm$ 42 }                                                 \\
\hline
Good primary vertex              &     \multicolumn{7}{c|}  { 2518 $\pm$ 42 }                                               \\
\hline
Two reconstructed \LJs              &  196    &   121    &   71     &    23   &    24   &   14     & 448 $\pm$ 11    \\
\hline
$\eta$ range (TYPE1/TYPE2)  &  196    &     83    &   32     &    13    &     9    &    5      & 337 $\pm$ 10   \\
\hline
EM fraction  (TYPE2)               &  196    &     83    &   11     &    13    &     6    &    1      &  308 $\pm$ 9  \\
\hline
Jet width W  (TYPE2)               &  196    &     83    &   11     &    13    &     6    &    1     &  308 $\pm$ 9  \\
\hline
Jet timing (TYPE1/TYPE2)      &   196    &     80    &   11     &    11   &     5     &   1     &  304 $\pm$ 9   \\
\hline
NC muons (TYPE0/TYPE1)      &   101    &     39    &    8     &     5    &      4    &   1     &  158 $\pm$ 6    \\
\hline
ID isolation                             &     72     &     24    &    6     &     3    &     2     &   1     &  107 $\pm$ 5    \\
\hline
\myabsdelphi                         & 70 $\pm$ 4 & 23 $\pm$ 2 &  5 $\pm$ 1 & 3 $\pm$ 1 & 2 $\pm$ 1 & $0^{+0.6}_{-0}$ &  104 $\pm$ 5 \\
\hline
\end{tabular}
\caption{Expected number of \LJ events for the four-\gammad FRVZ model, using the parameter values in table~\ref{tab:param}. The numbers refer to selected signal events at different stages of the selection process and for each \LJ pair type. The number of signal events is rescaled to the \intlumi total integrated luminosity and the quoted uncertainties are statistical only. The detection efficiency is 2.6 $\times 10^{-3}$.}
\label{tab:cutFlowFRVZ4}
\end{table}
%
%%%%%%%%%%%%%%%%%%%%%%%%% S E C T I O N %%%%%%%%%%%%%%%%%%%%%%%%%%%%%%%%%%%%%%%%%
\section{Systematic uncertainties}
\label{sec:Syst}
The following effects are considered as possible sources of systematic uncertainty and are included as input to obtain, using the {\it CLs} method~\cite{CLspaper}, the upper limits on the \sigmabr for the processes \higgstwogd and \higgsfourgd of the FRVZ models.
\begin{itemize}
\item{\bf Luminosity} \\
The overall normalization uncertainty of the integrated luminosity is 2.8$\%$. The systematic uncertainty on the luminosity is derived following the same methodology as that detailed in ref.~\cite{LUMI1}.
\item{\bf Higgs production cross section } \\
The uncertainty on the Higgs boson gluon fusion production cross section at $\sqrt{s}$~=~8~\mytev is 8$\%$ \cite{HiggsCrossS}.
\item{\bf Trigger} \\
The systematic uncertainty on the 3mu6 trigger efficiency was assumed to be
 dominated by the systematic on the 2mu6 trigger. The systematic uncertainty on the 2mu6 trigger efficiency was evaluated using a \tagandprobe~method applied to $J/\psi \to \mu\mu$ 2012 data and MC samples; it amounts to 5.8$\%$. The systematic uncertainty on the calorimetric trigger was evaluated for each requirement at L2 \cite{caloRatio}; the largest uncertainty, coming from the low EM fraction requirement, is 11$\%$.
\item{\bf Muon reconstruction efficiency} \\
The systematic uncertainty on the single \gammad reconstruction efficiency is evaluated using the \tagandprobe~method applied to $J/\psi \to \mu\mu$ 2012 data and MC samples. The $J/\psi\rightarrow \mu^{+}\mu^{-}$ decays were selected and the efficiency evaluated as a function of the opening angle $\Delta R$ between the two muons, both for data and for $J/\psi$ MC events (figure~\ref{fig:TP_rec_effi_data_MC} (top)). The two measures differ by less than two standard deviations at each point as shown in figure~\ref{fig:TP_rec_effi_data_MC} (bottom).
\begin{figure}[t!]
\centering
\includegraphics[width=80mm]{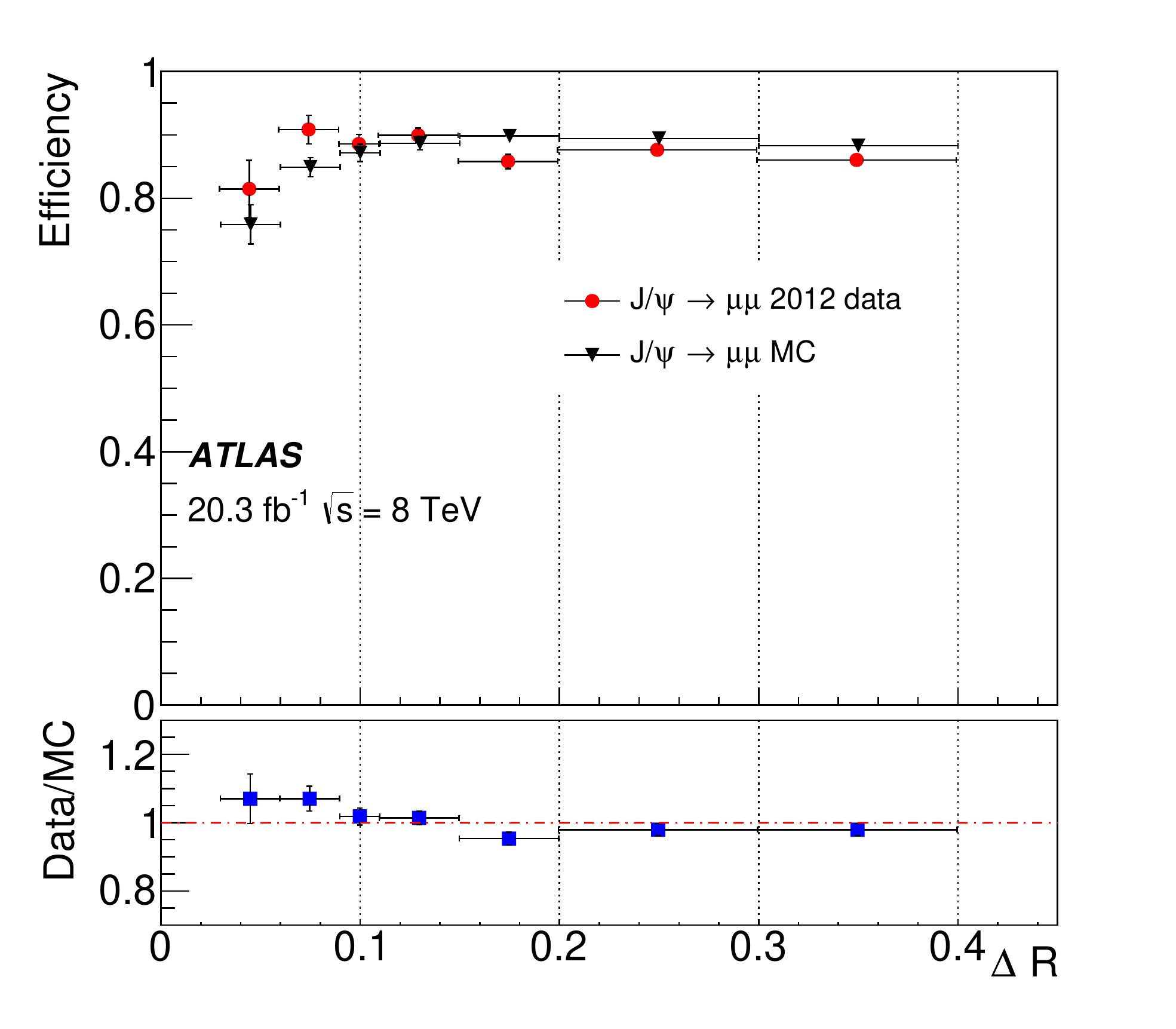}
\caption{Reconstruction efficiency of single NC muon with the \tagandprobe~method as a function of $\Delta R$ between the two muons in data and the $J/\psi\rightarrow \mu^{+}\mu^{-}$ MC sample (top), and the corresponding ratio of these two efficiencies (bottom).}
\label{fig:TP_rec_effi_data_MC}
\end{figure}
For low $\Delta R$ values the efficiency decreases because it is more difficult for the MS tracking algorithms to reconstruct two tracks with small angular separation. The resulting systematic uncertainty is 5.4$\%$.
\item{\bf Muon momentum resolution} \\
The systematic uncertainty on the muon momentum resolution for NC muons was evaluated by smearing and shifting the momentum of the muons by scale factors derived from comparison of \Zmumu~decays in data and MC simulation, and by observing the effect of this shift on the signal efficiency. The overall effect of the muon momentum resolution uncertainty is negligible.
\item{\bf Jet energy scale (JES)} \\
The effect of the JES uncertainty components~\cite{JetEnScale}~was evaluated for the jets of the TYPE1 and TYPE2 LJs. This uncertainty was applied to the MC signal samples. The variation in event yield amounts to 0.9$\%$ and to 1.7$\%$ for the two-\gammad and four-\gammad samples, respectively.
\item{\bf Effect of pile-up on \SumpT} \\
The presence of multiple collisions per bunch crossing (pile-up) affects the efficiency of the ID isolation criterion defined by the \SumpT~variable. The systematic uncertainty on the \SumpT~isolation efficiency due to pile-up is evaluated by computing the isolation efficiency $\varepsilon \left( \sum p_{\mathrm{T}} \right)$ for muons from  a sample of reconstructed $Z \to \mu\mu$ in data, as a function of the number of interaction vertices in the event.\footnote{The $\varepsilon \left( \sum p_{\mathrm{T}} \right)$ efficiency is defined as the number of muons with \SumpT not exceeding a given value, divided by the total number of muons. The ID track matched with the muon is removed from the sum.} The distributions of the isolation efficiency as a function of the isolation variable, for four subsamples of events with an increasing number of interaction vertices are shown in figure~\ref{fig:effi_isolation_pileup}. The effect of pile-up on the isolation efficiency is quantified by assuming for it the uniform distribution (worst case). The corresponding variance computed at \SumpT~=~3~\mygev was assumed as systematic uncertainty. It amounts to 4.1$\%$.
\begin{figure}[t!]
\centering
\includegraphics[width=80mm]{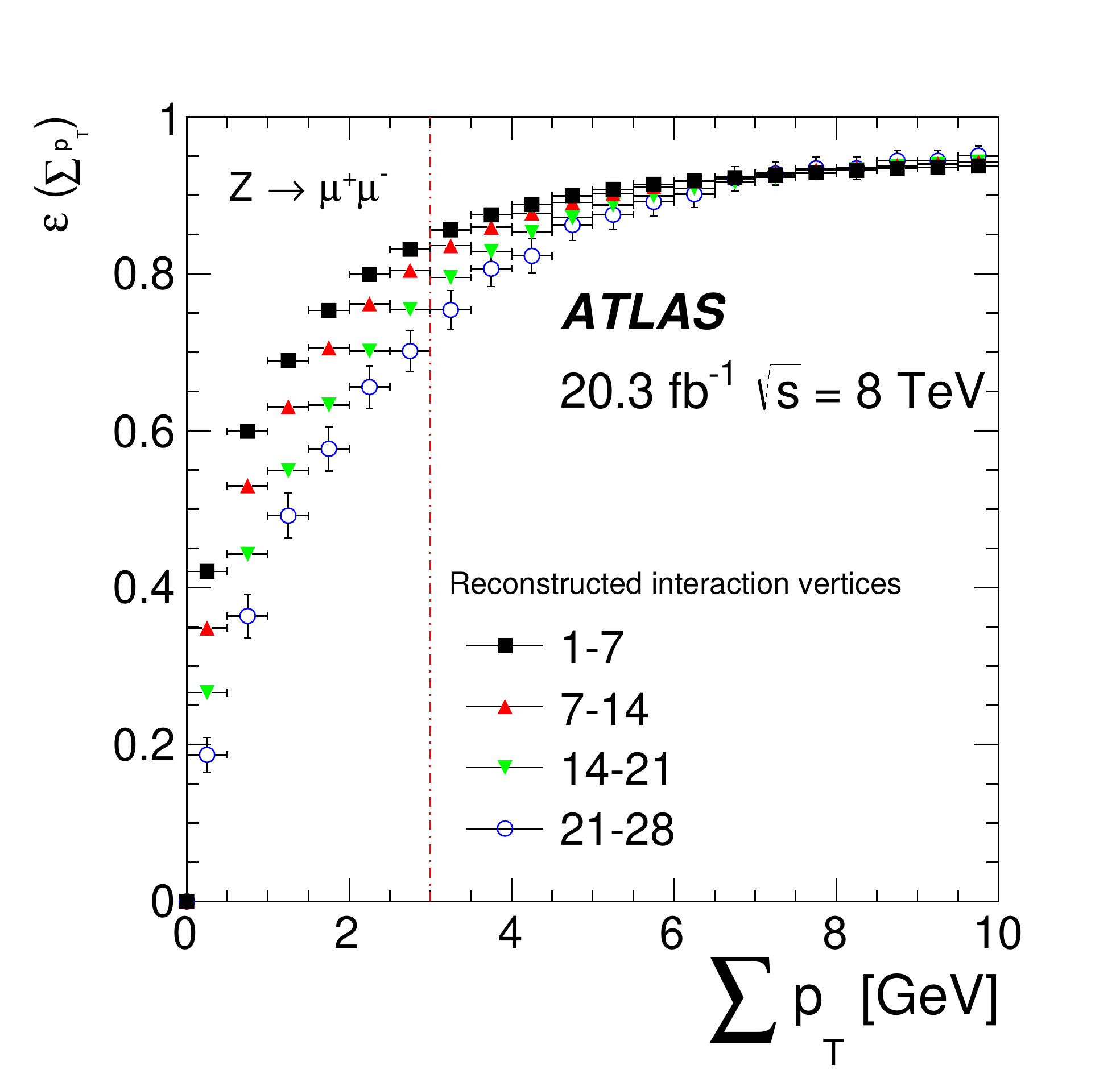}
\caption{Isolation efficiency as a function of \SumpT for four intervals of the number of reconstructed interaction vertices per event in a $Z \to \mu\mu$ data sample.}
\label{fig:effi_isolation_pileup}
\end{figure}
\item{\bf Multi-jet background estimation} \\
The systematic uncertainties that can affect the multi-jet background evaluation are related to the data-driven method used. The limits used to define the various regions were changed to take into account the expected uncertainty on \myabsdelphi~(comparing the \LJ direction at the MC generator level with the reconstructed direction, $\sigma_{|\Delta \phi|}=$~0.1~rad) and on $\sum{p_{\mathrm{T}}}$ (from the isolation distribution using the $Z \to \mu\mu$ data sample, $\sigma_{\sum{p_{\mathrm{T}}}}=$~0.6~\mygev). The background values were recomputed. This systematic uncertainty amounts to $15\%$. The additional effect of signal leakage into the control regions is taken into account by the simultaneous ABCD method used (see section~\ref{sec:results}).
\item{\bf Cosmic-ray background}\\
The systematic uncertainty on the cosmic-ray background is taken to be the statistical uncertainty on the number of cosmic-ray events in region D of the ABCD matrix (see table \ref{tab:abcd} and \ref{tab:abcd-no22}). The overall uncertainty is 22$\%$. Excluding the TYPE2-TYPE2 events it is 100$\%$.
\item{\bf \gammad detection efficiency and \pt~resolution} \\
Two additional effects were considered: the statistical uncertainty on the detection efficiency as a function of the decay position of the \gammad (see figures~\ref{fig:type0_recoeff_dp} and \ref{fig:type2_recoeff_dp}) and the resolution effects on the  \pt~of the \gammad. The reconstructed  \pt~of the \gammad differs from the MC generator-level \pt~value, inducing a 10$\%$ uncertainty.
\end{itemize}
%%%%%%%%%%%%%%%%%%%%%%%%% S E C T I O N %%%%%%%%%%%%%%%%%%%%%%%%%%%%%%%%%%%%%%%%%
\section{Results and interpretation}
\label{sec:results}
\begin{figure}[t!]
\centering
\includegraphics[width=80mm]{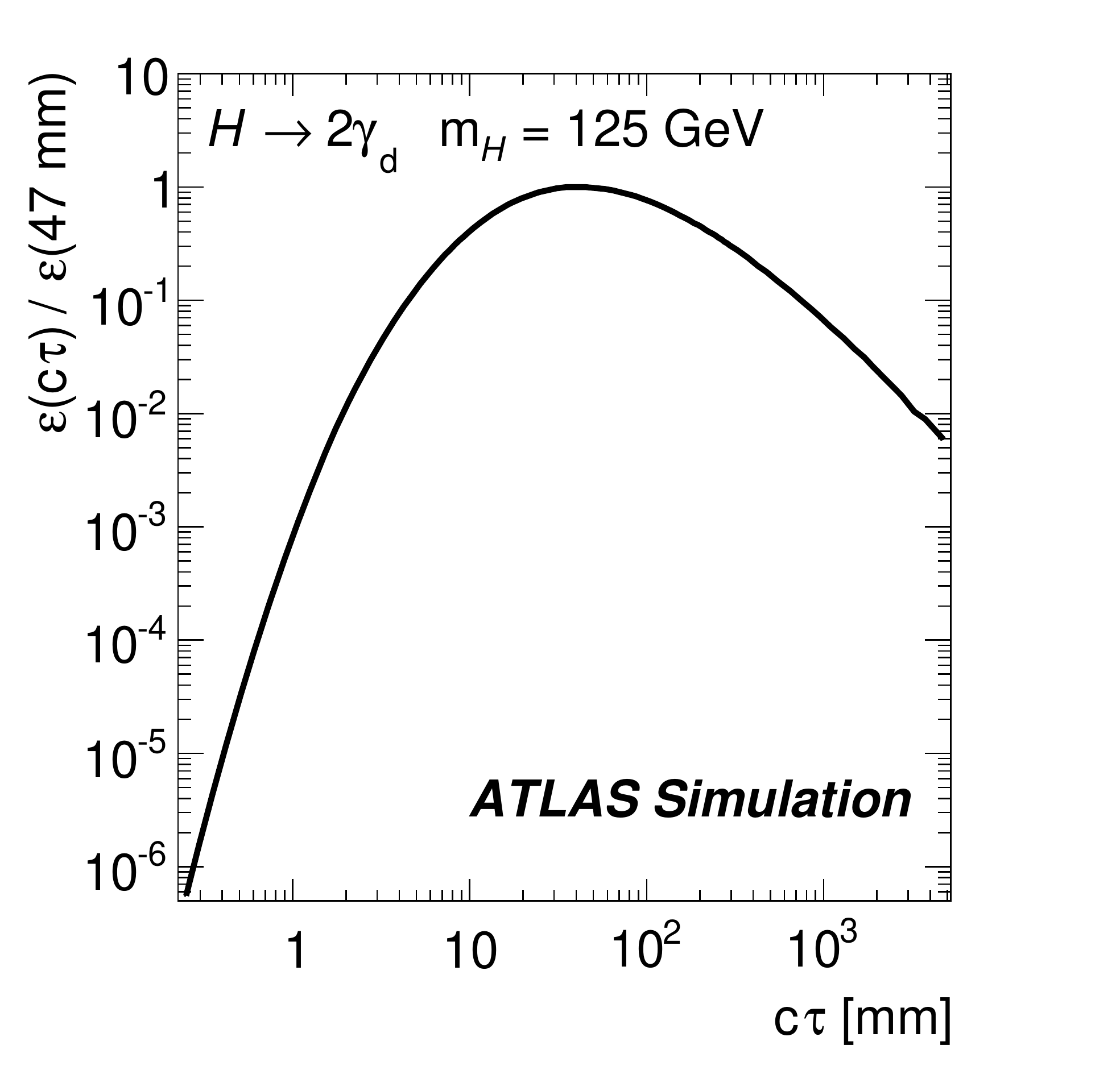}
\caption{Ratio of the integrated detection efficiency at a given \ctau~to the detection efficiency at \ctau=~47~mm of the reference \higgstwogd MC sample.}
\label{fig::rescale}
\end{figure}

\begin{figure}[t!]
\centering
\includegraphics[width=70mm]{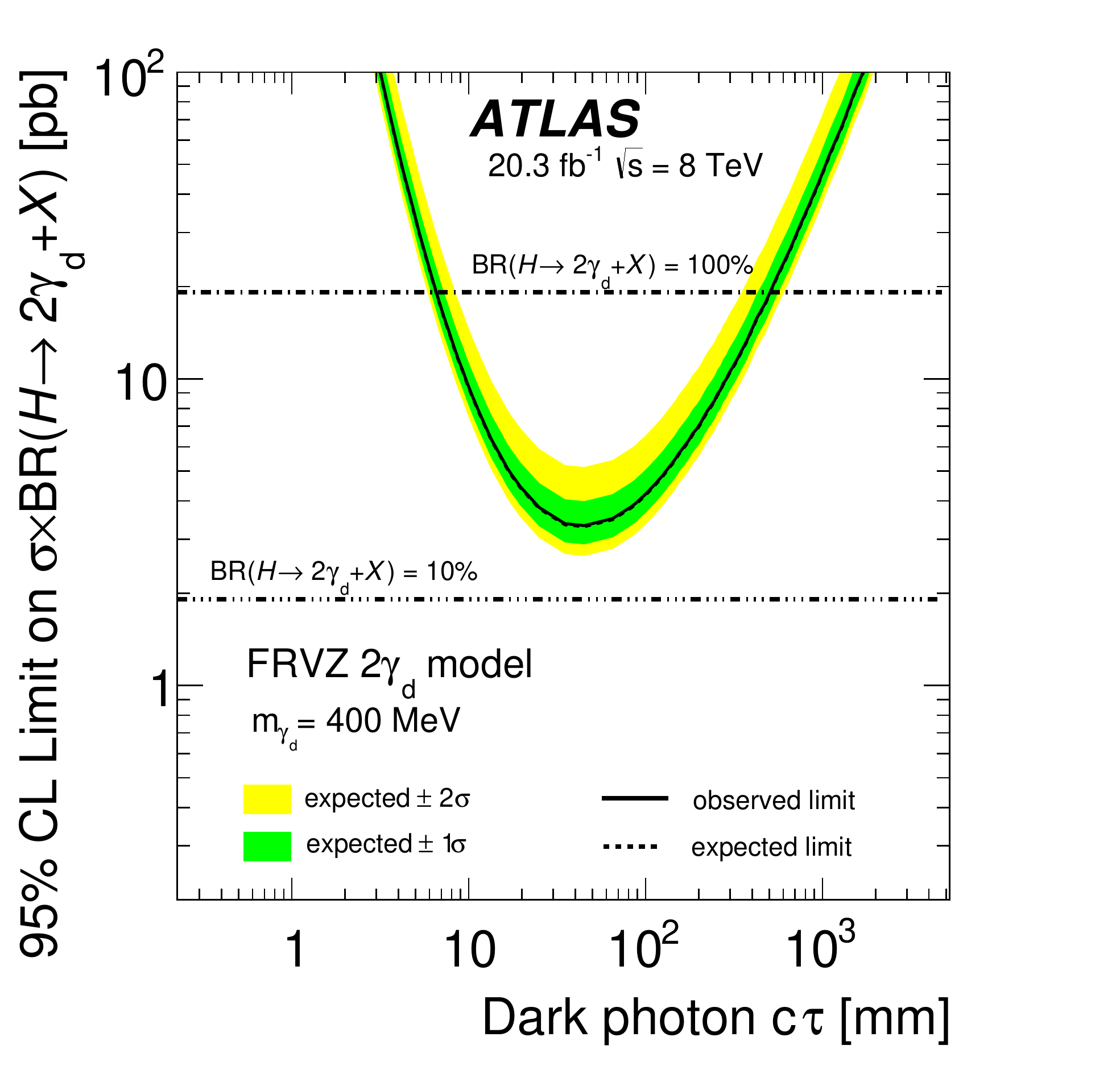}
\includegraphics[width=70mm]{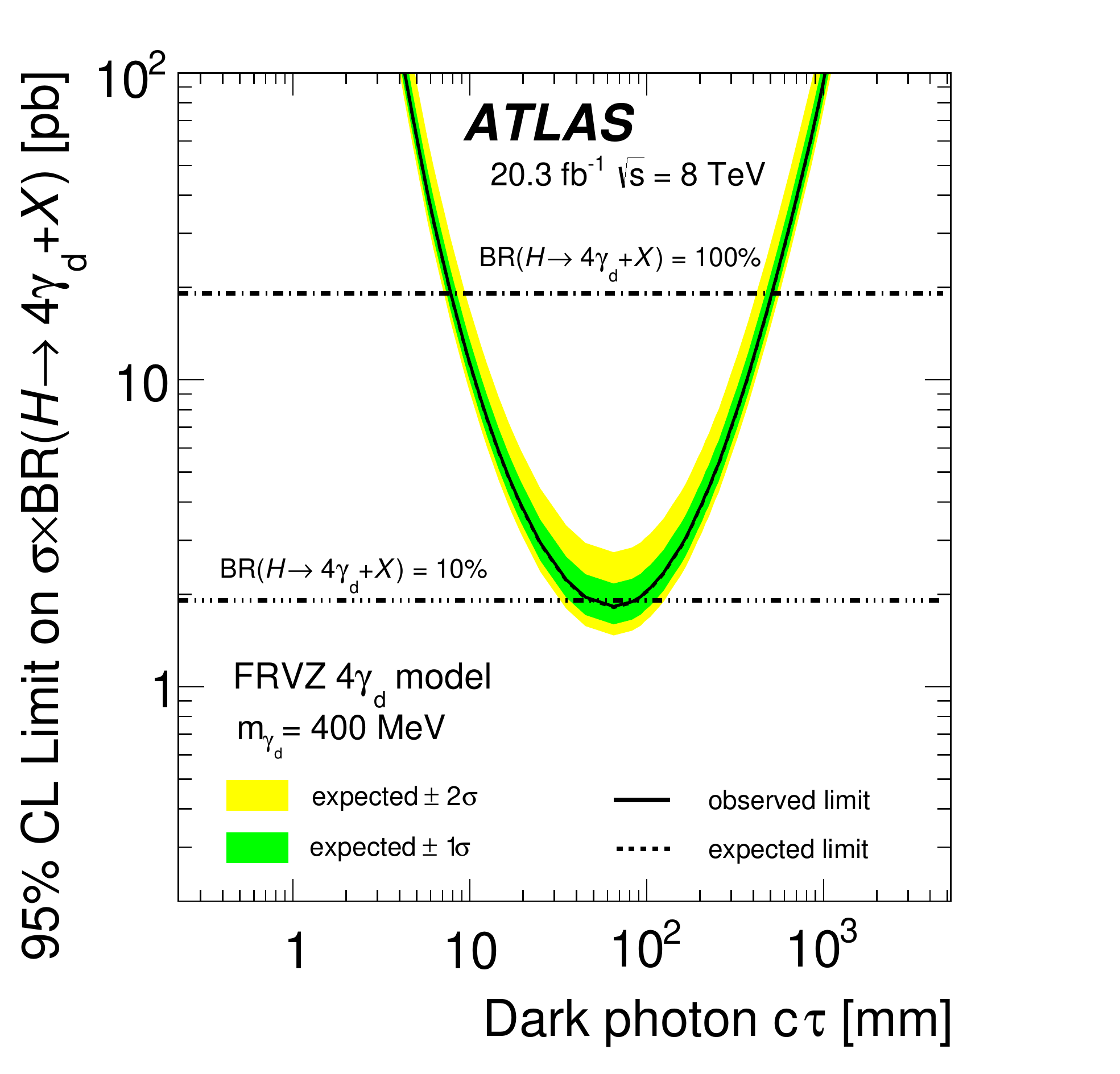}
\caption{The 95$\%$ upper limits on the \sigmabr for the processes \higgstwogd (left) and \higgsfourgd (right), as a function of the \gammad lifetime (\ctau) for the FRVZ benchmark samples.
The expected limit is shown as the dashed curve and the almost identical solid curve shows the observed limit. The horizontal lines correspond to \sigmabr for two values of the BR of the Higgs boson decay to \gammads.}
\label{fig:CLboth}
\end{figure}

\begin{figure}[t!]
\centering
\includegraphics[width=70mm]{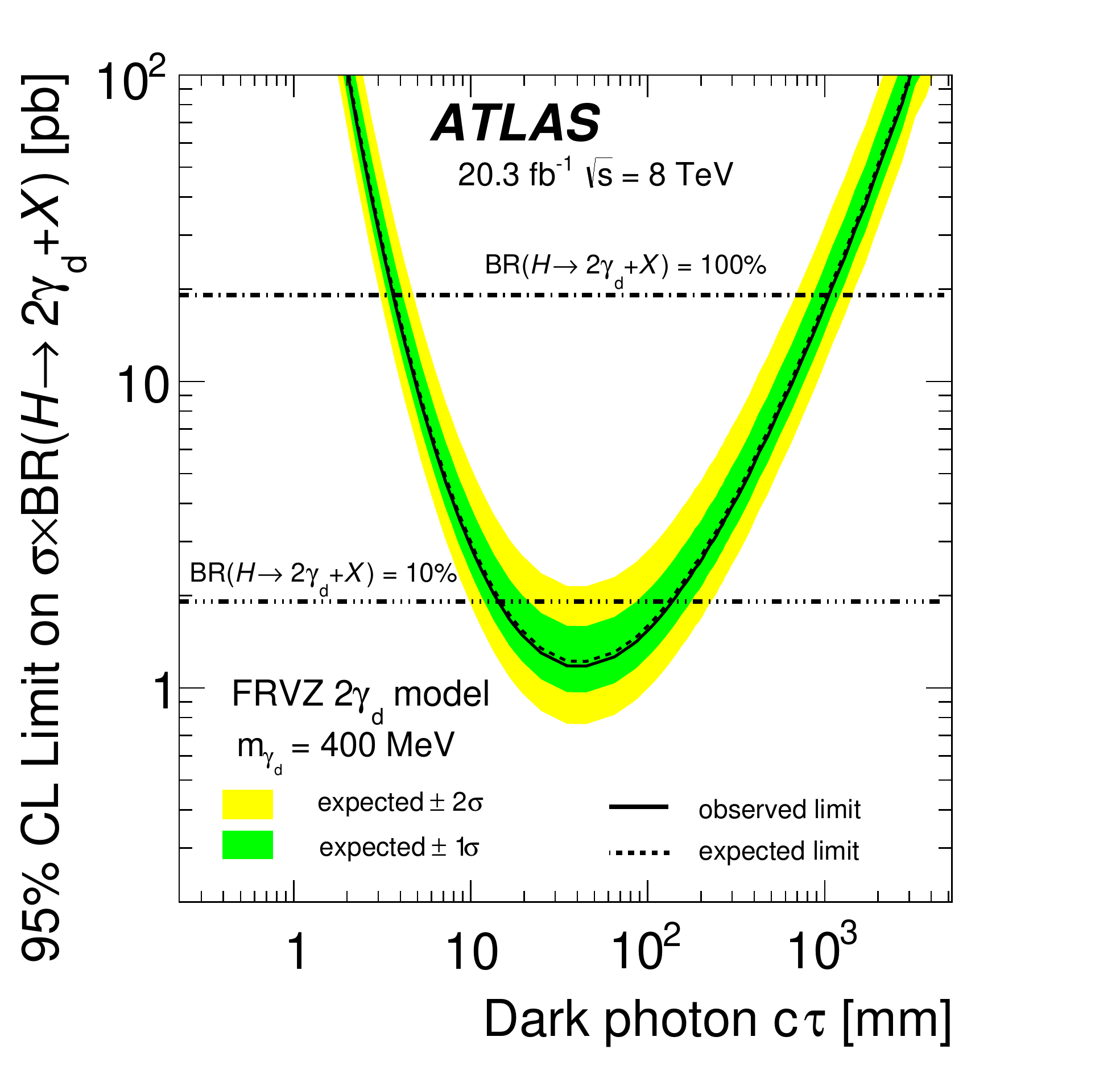}
\includegraphics[width=70mm]{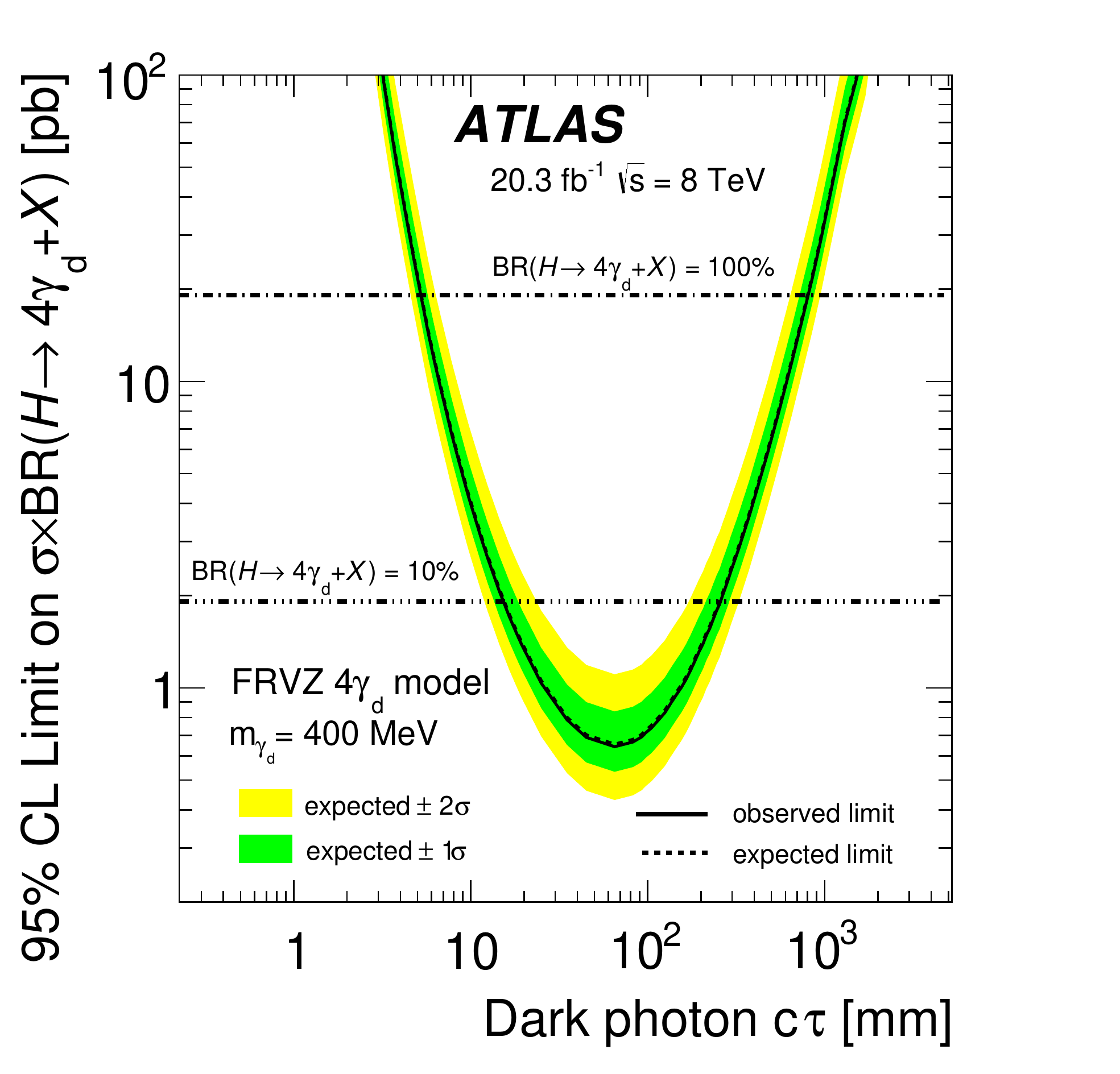}
\caption{The 95$\%$ upper limits on the \sigmabr for the processes \higgstwogd (left) and \higgsfourgd (right), as a function of the \gammad lifetime (\ctau) for the FRVZ benchmark samples, excluding the TYPE2-TYPE2 events. The expected limit is shown as the dashed curve and the almost identical solid curve shows the observed limit.  The horizontal lines correspond to \sigmabr for two values of the BR of the Higgs boson decay to \gammads.}
\label{fig:CLboth-no22}
\end{figure}

\begin{figure}[ht!]
\centering
\includegraphics[width=150mm]{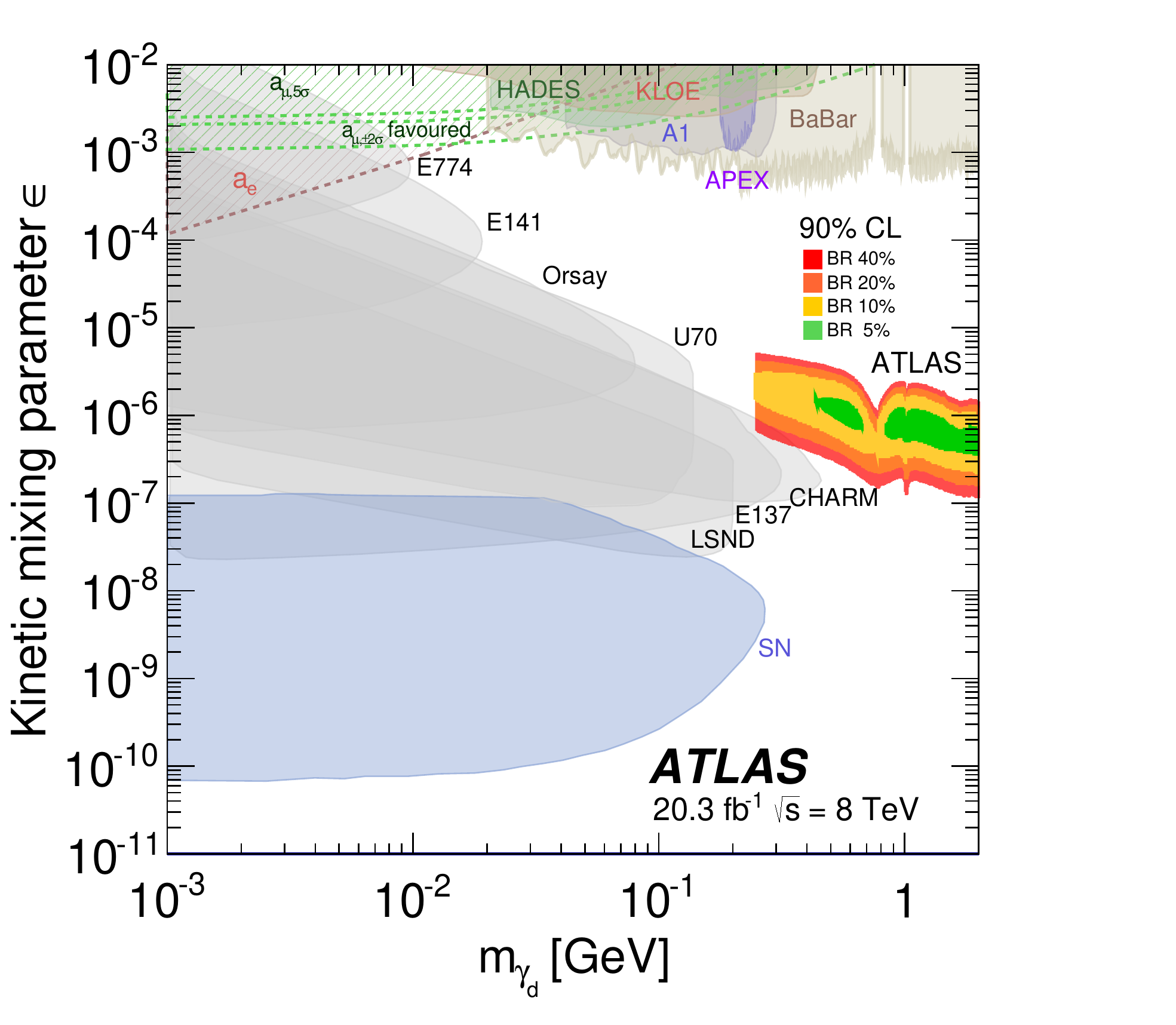}
\caption{ Parameter space  exclusion plot for dark photons as a function of the \gammad mass and of the kinetic mixing parameter $\epsilon$ from figure~6 of ref.~\cite{snowmass}. Shown are existing 90$\%$ CL exclusion regions from beam dump experiments E137, E141, and E774~\cite{117,118,119}, Orsay~\cite{Orsay}, U70~\cite{U70}, CHARM~\cite{CHARM}, LSND~\cite{LSND}, A1~\cite{127}, the electron and muon anomalous magnetic moment \cite{120,121,122}, HADES~\cite{Hades}, KLOE~\cite{123,124}, the test run results reported by APEX~\cite{126}, an estimate using a BaBar result~\cite{128,129,132}, and constraints from astrophysical observations~\cite{130,131}. The 90$\%$ CL exclusion limits from the present search, assuming the FRVZ model \higgstwogd with decay branching fraction to \gammad of 5/10/20/40$\%$  and the NNLO gluon fusion Higgs production cross section, are shown. }
\label{fig:DMexclusion}
\end{figure}

Table~\ref{tab:resultsfinale} summarizes the data and background results of the search for LJs in the 2012 data sample. Both for all LJ pair events and for the case where the TYPE2-TYPE2 LJs are excluded the data agree with the background expectation.
\begin{table}[ht!]
\centering
\small
\begin{tabular}{| c | c | c |}
\hline
                                  & All LJ pair types & TYPE2-TYPE2 \LJs excluded \\
\hline
\hline
       Data                    &        119            &                 29                        \\
\hline
      Cosmic rays           &   40 $\pm$ 11  $\pm$ 9  &           29 $\pm$ 9  $\pm$ 29              \\
\hline
     Multi-jets (ABCD)   &   70 $\pm$ 58 $\pm$ 11  &           12 $\pm$ 9 $\pm$ 2                \\
\hline
     Total background  & 110 $\pm$ 59 $\pm$ 14  &           41 $\pm$ 12 $\pm$ 29              \\
\hline
\end{tabular}
\caption{Summary of the \LJ selection applied to data and background in the full 2012 data sample. The first uncertainty is statistical, while the second is systematic.}
\label{tab:resultsfinale}
\end{table}
\\
The results of the search for \LJ production are used to set upper limits on the Higgs boson decay branching fraction to \LJs as a function of the \gammad mean lifetime, according to the FRVZ models. The efficiency of the selection criteria described above is evaluated for the simulated FRVZ model samples  as a function of the mean lifetime of the \gammad. The signal MC events are weighted by the detection probability of the \gammad in the various regions of the detector, generating their decay points according to a chosen value of the \gammad proper decay length (\ctau times the \gammad Lorentz factor), with \ctau~ranging from 0.5 to 4750~mm. The number of selected events are then rescaled by the ratio of the integrated detection efficiency at a given \ctau, $\rm \varepsilon(c\tau)$, to the efficiency for the reference sample, $\rm \varepsilon(47~\rm{mm} )$ (see table \ref{tab:cutFlowFRVZ2} and \ref{tab:cutFlowFRVZ4}). Figure~\ref{fig::rescale} shows, for the \higgstwogd model, the ratio $\rm {\varepsilon(c\tau)} /{\varepsilon(47~\rm{mm} )}$ as a function of \ctau. These numbers, together with the expected number of background events (multi-jets and cosmic rays), are used as input to obtain limits at the 95$\%$ confidence level (CL) on the cross section times branching ratio (\sigmabr) for the processes \higgstwogd and \higgsfourgd.
The simultaneous {\it CLs} method is used to determine the limits, where the ABCD regions are populated from the data-driven background estimate and from the appropriate signal hypothesis. It also takes into account contaminations from sources of background other than QCD processes. \\
All the systematic uncertainties discussed in section~\ref{sec:Syst}, except the ones on the signal MC cross sections, and their correlations are taken into account in calculating the limits. As a final cross-check the number of expected multi-jet background events in the signal region from the simultaneous {\it CLs} ABCD method, can be compared with the expected background from the ABCD method assuming no signal (see section~\ref{sec:ABCDQCD}). For the two-\gammad model the estimated background is 13~$\pm$~8 events and for the four-\gammad model it is 13~$\pm$~7 events, to be compared with 12~$\pm$~9 events obtained by ABCD method assuming no signal (section~\ref{sec:ABCDQCD}). The resulting exclusion limits on the \sigmabr, assuming the Higgs boson SM gluon fusion production cross section $\sigma_{\textrm\SMs} = $ 19.2~pb, are shown in figure~\ref{fig:CLboth} as a function of the \gammad mean lifetime (expressed as \ctau) for the two models.
The exclusion plots with the TYPE2-TYPE2 category of events removed are shown in figure~\ref{fig:CLboth-no22}.
In figure~\ref{fig:CLboth} and figure~\ref{fig:CLboth-no22} the observed limit is slightly better than the expected one, because the number of events in the signal region is slightly smaller than the expected background from cosmic rays and multi-jets. It is seen that for these two models the search is more sensitive when excluding the TYPE2-TYPE2 events. Table~\ref{tab:limits-22} shows the ranges in which the \gammad lifetime (\ctau) is excluded at the 95$\%$ CL for \higgstwogd and \higgsfourgd assuming a BR of 10$\%$. The corresponding limits with TYPE2-TYPE2 events excluded are shown in table~\ref{tab:limits}. \\
For the case of a hidden photon which kinetically mixes with the SM photon, these limits can be converted into exclusion limits on the kinetic mixing parameter $\epsilon$ using the eqs. (4) and (5) of ref.~\cite{epsilon}. For more details see also refs.~\cite{b2, b10b}. For \higgstwogd with a \gammad  mass = 0.4~\mygev excluding TYPE2-TYPE2 events, the interval that is excluded at 95$\%$ CL is 7.7$\times10^{-7}$~$\leq$~$\epsilon$~$\leq$~2.7$\times10^{-6}$.\\
These results are also interpreted in the context of the Vector portal model as exclusion contours in the kinetic mixing parameter $\epsilon$ vs \gammad mass plane~\cite{snowmass,116} as shown in figure~\ref{fig:DMexclusion}. Assuming Higgs decay branching fractions into \gammad of 5/10/20/40$\%$ and the NNLO gluon fusion Higgs production cross section, the lifetime limits can be converted into kinetic mixing parameter $\epsilon$ limits. While the other limits are model-independent because they produce the hidden photon through the vector portal coupling, this limit does depend on the additional assumption on the Higgs branching fraction to the hidden sector. The resulting 90$\%$ CL exclusion regions for \higgstwogd are shown in figure~\ref{fig:DMexclusion}; the \gammad mass interval ($0.25\mbox{--}1.5$)~\mygev corresponds to the values in which the \gammad decay branching fractions and the detection efficiencies are comparable with those for the 0.4~\mygev \gammad mass. The systematic uncertainties due to the detection efficiency and  decay branching fraction variations as a function of the \gammad mass were estimated and included in the 90$\%$ CL exclusion region evaluations.
\renewcommand{\arraystretch}{1.2}
\begin{table}[t!]
\small
\centering
\begin{tabular}{|c|c|}\hline
     FRVZ model          &        Excluded c$\tau$ $[\rm{mm} ]$  \\
                                 &                 BR(10$\%$)                \\
\hline
\hline
    \higgstwogd         &                  no limit                     \\
\hline
    \higgsfourgd        &    52 $\leq$ c$\tau$ $\leq$ 85  \\
\hline
\end{tabular}
\caption{Ranges of \gammad lifetime (\ctau) excluded at 95$\%$ CL for \higgstwogd  and \higgsfourgd, assuming 10$\%$ BR and the Higgs boson SM gluon fusion production cross section and including the TYPE2-TYPE2 events.}
\label{tab:limits-22}
\end{table}
\renewcommand{\arraystretch}{1.2}
\begin{table}[t!]
\small
\centering
\begin{tabular}{|c|c|}\hline
      FRVZ model           &         Excluded c$\tau$ $[\rm{mm} ]$    \\
                                   &                 BR(10$\%$)                   \\
\hline
\hline
     \higgstwogd          &    14 $\leq$ c$\tau$ $\leq$ 140   \\
\hline
     \higgsfourgd         &    15 $\leq$ c$\tau$ $\leq$ 260    \\
\hline
\end{tabular}
\caption{Ranges of \gammad lifetime (\ctau) excluded at 95$\%$ CL for \higgstwogd  and \higgsfourgd, assuming  10$\%$ BR and the Higgs boson SM gluon fusion production cross section. TYPE2-TYPE2 events are not used.}
\label{tab:limits}
\end{table}
%
%\clearpage
%%%%%%%%%%%%%%%%%%%%%%%%% S E C T I O N %%%%%%%%%%%%%%%%%%%%%%%%%%%%%%%%%%%%%%%%
%
%

%%%%%%%%%%%%%%%%%%%%%%%%% S E C T I O N %%%%%%%%%%%%%%%%%%%%%%%%%%%%%%%%%%%%%%%%
\section{Conclusions}
\label{sec:Conclusions}
The ATLAS detector at the LHC is used to search for the production of non-prompt \LJs in a \intlumi sample of $\sqrt{s} =8$~\mytev~{\it pp} collisions. Starting from a quite general definition of non-prompt \LJs, a set of selection criteria able to isolate their signature from the SM and cosmic-rays backgrounds were defined. The observed data are consistent with the experimental background expectations. This result can be used to set upper limits on non-SM Higgs boson decays to \LJs according to the FRVZ models. Limits are set on the \sigmabr for \higgstwogd and \higgsfourgd  for $m_{H}=$ 125~\mygev and a \gammad mass of 0.4~\mygev, as a function of the long-lived particle mean lifetime. Assuming the SM gluon fusion production cross section for a 125~\mygev Higgs boson, its BR to hidden-sector photons is found to be below 10$\%$, at 95$\%$ CL, for hidden photon \ctau in the range 14~mm~$\leq$~c$\tau$~$\leq$~140~mm for the \higgstwogd model and in the range 15~mm~$\leq$~c$\tau$~$\leq$~260~mm for the \higgsfourgd model. \\
These results are also interpreted in the context of the Vector portal model as exclusion contours in the kinetic mixing parameter $\epsilon$ vs \gammad mass plane and significantly improve the constraints from other experiments.
%%%%%%%%%%%%%%%%%%%%%%%%%%%%%%%%%%%%%%%%%%%%%%%%%%%%%%%%%%%%%%%%
%
\section*{Acknowledgements}
We thank CERN for the very successful operation of the LHC, as well as the
support staff from our institutions without whom ATLAS could not be
operated efficiently.
We acknowledge the support of ANPCyT, Argentina; YerPhI, Armenia; ARC,
Australia; BMWFW and FWF, Austria; ANAS, Azerbaijan; SSTC, Belarus; CNPq and FAPESP,
Brazil; NSERC, NRC and CFI, Canada; CERN; CONICYT, Chile; CAS, MOST and NSFC,
China; COLCIENCIAS, Colombia; MSMT CR, MPO CR and VSC CR, Czech Republic;
DNRF, DNSRC and Lundbeck Foundation, Denmark; EPLANET, ERC and NSRF, European Union;
IN2P3-CNRS, CEA-DSM/IRFU, France; GNSF, Georgia; BMBF, DFG, HGF, MPG and AvH
Foundation, Germany; GSRT and NSRF, Greece; ISF, MINERVA, GIF, I-CORE and Benoziyo Center,
Israel; INFN, Italy; MEXT and JSPS, Japan; CNRST, Morocco; FOM and NWO,
Netherlands; BRF and RCN, Norway; MNiSW and NCN, Poland; GRICES and FCT, Portugal; MNE/IFA, Romania; MES of Russia and ROSATOM, Russian Federation; JINR; MSTD,
Serbia; MSSR, Slovakia; ARRS and MIZ\v{S}, Slovenia; DST/NRF, South Africa;
MINECO, Spain; SRC and Wallenberg Foundation, Sweden; SER, SNSF and Cantons of
Bern and Geneva, Switzerland; NSC, Taiwan; TAEK, Turkey; STFC, the Royal
Society and Leverhulme Trust, United Kingdom; DOE and NSF, United States of
America.
The crucial computing support from all WLCG partners is acknowledged
gratefully, in particular from CERN and the ATLAS Tier-1 facilities at
TRIUMF (Canada), NDGF (Denmark, Norway, Sweden), CC-IN2P3 (France),
KIT/GridKA (Germany), INFN-CNAF (Italy), NL-T1 (Netherlands), PIC (Spain),
ASGC (Taiwan), RAL (UK) and BNL (USA) and in the Tier-2 facilities
worldwide.
%
%%%%%%%%%%%%%%%%%%%%%%%%%%%%%%%%%%%%%%%%%%%%%%%%%%%%%%%%%%%%%%%%

%
% ATLAS Collaboration author list
% Data extracted on 21-Nov-2014 for paper reference EXOT-2013-22
%\documentclass[11pt]{article}
%\usepackage{a4wide}\begin{document}
\begin{flushleft}
{\Large The ATLAS Collaboration}

\bigskip

G.~Aad$^{\rm 85}$,
B.~Abbott$^{\rm 113}$,
J.~Abdallah$^{\rm 153}$,
S.~Abdel~Khalek$^{\rm 117}$,
O.~Abdinov$^{\rm 11}$,
R.~Aben$^{\rm 107}$,
B.~Abi$^{\rm 114}$,
M.~Abolins$^{\rm 90}$,
O.S.~AbouZeid$^{\rm 160}$,
H.~Abramowicz$^{\rm 155}$
% ...
\bigskip
\\
% ...
$^{11}$ Institute of Physics, Azerbaijan Academy of Sciences, Baku, Azerbaijan\\
% ...
$^{85}$ CPPM, Aix-Marseille Universit{\'e} and CNRS/IN2P3, Marseille, France\\
% ...
$^{90}$ Department of Physics and Astronomy, Michigan State University, East Lansing MI, United States of America\\
% ...
$^{107}$ Nikhef National Institute for Subatomic Physics and University of Amsterdam, Amsterdam, Netherlands\\
% ...
$^{113}$ Homer L. Dodge Department of Physics and Astronomy, University of Oklahoma, Norman OK, United States of America\\
$^{114}$ Department of Physics, Oklahoma State University, Stillwater OK, United States of America\\
% ...
$^{117}$ LAL, Universit{\'e} Paris-Sud and CNRS/IN2P3, Orsay, France\\
% ...
$^{153}$ Institute of Physics, Academia Sinica, Taipei, Taiwan\\
% ...
$^{155}$ Raymond and Beverly Sackler School of Physics and Astronomy, Tel Aviv University, Tel Aviv, Israel\\
% ...
$^{160}$ Department of Physics, University of Toronto, Toronto ON, Canada\\
% ...
\end{flushleft}

%\end{document}
% Created with ./xml2latex.py 
%
\clearpage
\end{document}